\documentclass[11pt]{article}

\usepackage{framed,url}
\usepackage{pstricks}
\usepackage{multicol}
\usepackage{graphicx}
\usepackage{graphics}
\usepackage{amsmath,latexsym,amssymb}
\usepackage{color}
\usepackage{subfigure}
\usepackage{rotate}
\usepackage{fancyheadings}
\usepackage{portland}
\usepackage{pifont}

\newcommand{\rem}[1]{}

\DeclareMathAlphabet{\mathbi}{OML}{cmm}{b}{it} 

\newtheorem{theorem}{Theorem}
\newtheorem{lemma}{Lemma}

\newcommand{\bx}{\mathbi{x}}

\newcommand{\bel}{\begin{equation}\label}
\newcommand{\ee}{\end{equation}}
\newcommand{\ben}{\begin{enumerate}}
\newcommand{\een}{\end{enumerate}}
\newcommand{\bit}{\begin{itemize}}
\newcommand{\eit}{\end{itemize}}
\newcommand{\I}{\int_{\mathcal{V}}}

\newcommand{\bfv}{\mathfrak{v}}

\newcommand{\bu}{\mathbi{u}}
\newcommand{\bv}{\mathbi{v}}

\newcommand{\bV}{\mathbi{V}}

\newcommand{\bom}{\mbox{\boldmath$\omega$}}

\newcommand{\bvom}{\mbox{\boldmath$\zeta$}}
\newcommand{\bvarpi}{\mbox{\boldmath$\mathfrak{w}$}}

\newcommand{\bcapom}{\mbox{\boldmath$\Omega$}}

\newcommand{\bzeta}{\mbox{\boldmath$\zeta$}}

\newcommand{\bivec}{\mbox{\boldmath$i$}}
\newcommand{\bjvec}{\mbox{\boldmath$j$}}
\newcommand{\bk}{\mbox{\boldmath$k$}}

\newcommand{\bn}{\mbox{\boldmath$\hat{n}$}}

\newcommand{\asp}{\alpha_{a}}
\newcommand{\beq}{\begin{eqnarray}\label} 
\newcommand{\eeq}{\end{eqnarray}} 
\newcommand{\bc}{\begin{center}} 
\newcommand{\ec}{\end{center}} 
\newcommand\shalf{\ensuremath{{\scriptstyle\frac{1}{2}}}}

\newcommand\threequart{\ensuremath{{\scriptstyle\frac{3}{4}}}}

\newcommand{\Rey}{Re}

\newcommand{\non}{\nonumber}
\newcommand{\varep}{\varepsilon}
\makeatletter\@addtoreset{equation}{section}\makeatother

\begin{document}
\sf
\title{\textbf{\sf Extreme events in solutions of hydrostatic and non-hydrostatic climate models}}
\author{
\hspace{-1.1cm}
J. D. Gibbon and D. D. Holm\\
Department of Mathematics, Imperial College London, London SW7 2AZ, UK}
\date{}

\maketitle
\makeatother
\maketitle
\tableofcontents

\begin{abstract}
Initially this paper reviews the mathematical issues surrounding the hydrostatic (HPE) and 
non-hydrostatic (NPE) primitive equations that have been used extensively in numerical weather 
prediction and climate modelling. Cao and Titi (2005, 2007) have provided a new impetus to 
this by proving existence and uniqueness of solutions of viscous HPE on a cylinder with 
Neumann-like boundary conditions on the top and bottom. In contrast, the regularity of solutions 
of NPE remains an open question. With this HPE regularity result in mind, the second issue 
examined in this paper is whether extreme events are allowed to arise spontaneously in their 
solutions. Such events could include, for example, the sudden appearance and disappearance of 
locally intense fronts that do not involve deep convection. Analytical methods are used to 
show that for viscous HPE, the creation of small-scale structures is allowed locally in space 
and time at sizes that scale inversely with the Reynolds number. 
\end{abstract}
\vspace{10mm}
\bc
A review dedicated to the memory of Robin Bullough.
\ec

\newpage

\section{\sf\textbf{Review of the hydrostatic \& non-hydrostatic primitive equations}}\label{intro}


The hydrostatic primitive equations (HPE) were developed more than eighty years ago by Richardson 
as a model for large-scale oceanographic and atmospheric dynamics \cite{Rich22}. In various forms 
they have been the foundation of most numerical weather, climate and global ocean circulation 
predictions for many decades. The HPE govern incompressible, rotating, stratified fluid flows that 
are in hydrostatic balance. This balance is broken, however, in \textit{deep convection} which 
occurs in cloud formation, flows over mountains and vertical fluid entrainment by strong gravity 
currents\footnote{\sf The subtle difference is explained in \S\ref{prim} between what is known as 
modellers' HPE (MHPE) pioneered by Richardson and incorporated in many numerical weather prediction 
and climate simulation models, and HPE in cartesian geometry discussed here.}. In this context, the 
\textit{hydrostatic approximation} is the most reliable and accurate of all the assumptions made 
in applying them to operational numerical weather prediction (NWP), climate modelling and global 
numerical simulations of ocean circulation. The approximation arises from a scale analysis of 
synoptic systems\,; see 
\cite{Charn55,Eck60,Gill82,Ped87,Salmon88,Ho92,Cullen93,Holm96,Marshall97,NR02,White02,
White03,DaCuMaMaStWhWo05,WhHoRoSt05,Cullen06,Cullen07,WS09}. 
Basically, it neglects the vertical acceleration term in the vertical momentum equation. Physically, 
this enforces a perfect vertical force balance between gravity and the vertical pressure gradient 
force, which means that the pressure at a given location is given by the weight of the fluid above. 

Until recently, the resolution of operational NWP models has been limited by computing power and 
operational time constraints on resolutions in which the hydrostatic approximation is almost 
perfectly valid. Operational forecasts therefore have relied mainly on hydrostatic models, which 
applied extremely well in numerical simulations of the global circulation of both the atmosphere 
and the ocean at the spatial and temporal resolutions that were available at the time. In parallel,
the development of non-hydrostatic atmospheric models, which retain the vertical acceleration term 
and thus capture strong vertical convection, has also been pursued over a period of more than four 
decades, especially for mesoscale investigations of sudden storms. As  computer systems have become 
faster and memory has become more affordable over the past decade, there has been a corresponding 
increase in the spatial resolution of both NWP and climate simulation models. This improvement has 
facilitated a transition from highly developed hydrostatic models towards non-hydrostatic models. 
Over the past decade, atmospheric research institutions have begun replacing their operational 
hydrostatic models with non-hydrostatic versions \cite{DaCuMaMaStWhWo05,WhHoRoSt05,Cullen07}. 

An impending `model upgrade' may be tested by comparing the solution properties of the existing 
hydrostatic model at a new higher resolution with those of its non-hydrostatic alternative (NPE). 
Using the fastest supercomputers available, global simulations of atmospheric circulation at 
resolutions of about 10 km in the horizontal (in the region of the hydrostatic limit) have recently been 
performed \cite{OhSaMaNa05,SACRLLC06}.  In the past these simulations have been carried out using 
only hydrostatic models, as the non-hydrostatic versions were still under development. Perhaps not 
unexpectedly, the simulations at finer resolution found much more fine-scale structure than had 
been seen previously. More importantly, the smallest coherent features found at the previous coarser 
limits of resolution were no longer present at new finer resolutions, because the balance 
between nonlinearity and dissipation that had created them previously was no longer being enforced. 
Instead, it was being enforced at the new limits of resolution.

Consequently, an important conclusion from these PE simulations has been that the transition from the 
existing hydrostatic model to its non-hydrostatic alternative, and the distinction between their 
solutions, will require in both cases a much better understanding of the formation of fine-scale 
frontal structures \cite{SACRLLC06,Ham08}. For discussions of the history of ideas in fronto-genesis 
see \cite{El48,H75,HB72,Hrev82}. An example of fine-scale wind measurements is a `gust front' are 
found in Figure 1 \cite{Cullen06,Shapiro84,gust1}. Gust fronts are indicated by `bow echoes' on D\"oppler 
radar. These have been extensively observed and studied as the precursors of severe local weather.

In particular, the physical parameterizations in hydrostatic models are likely to require dramatic 
changes when simulated at new finer resolutions. Therefore, in order to obtain the full benefit 
of the potential increase in accuracy provided by such high spatial resolution when using the 
non-hydrostatic model, one must first determine the range of solution behavior arising at the finer 
resolution in the computations of the previous hydrostatic model \cite{SACRLLC06}. Over relatively 
small domains, operational model resolutions in the atmosphere are in fact already beyond the hydrostatic 
limit. These models run at resolutions finer than \textit{10 km}, where convection is partially resolved 
(thus, inaccurately simulated) by non-hydrostatic models. However, convection can only be fully resolved 
at model resolutions of about two orders of magnitude finer, that is, about \textit{100 m} or less in the 
horizontal. It is therefore likely that, for many years to come, operational hydrostatic and non-hydrostatic 
models will function at resolutions where convection can only be partially resolved. Consequently, the 
distinction between hydrostatic and non-hydrostatic solution behavior at these intermediate resolutions 
becomes paramount for the accuracy, predictability, reliability and physical interpretation of results 
in NWP. 

\begin{figure}[ht!]
\centering
\includegraphics[width=.8\textwidth,angle=0]{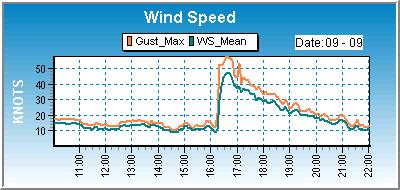}
\caption{\small\sf An example of a gust front manifesting itself in a sudden rise in wind 
speed \cite{Cullen06,Shapiro84}.}\label{front-fig}
\end{figure}

\par\vspace{-2mm}\noindent
Against this background of renewed interest in the improvement of hydrostatic models and their newly 
acquired computability at higher resolution, this paper also reviews the development of new mathematical 
ideas introduced by Cao and Titi \cite{CT05,CT07}, who have proved the existence and uniqueness of 
\textit{strong} solutions of the viscous hydrostatic primitive equations\footnote{\sf See also Kobelkov 
\cite{Kob06,Kob07a,Kob07b} for a subsequent independent method.}. The corresponding regularity problem 
for viscous NPE remains open\footnote{\sf This is closely related to the open regularity problem for 
the three-dimensional Navier-Stokes equations which has been solved only for very thin domains ($0 < 
\asp \ll 1$)\,: see \cite{thin1,thin2,thin3}.}.  Ju \cite{NJ07} then built on the result of Cao and 
Titi to show that the viscous HPE actually possess a global attractor, which means that the bounds 
are constants although no explicit estimates were given. The status of results prior to the Cao and 
Titi proof \cite{CT05,CT07} (e.g. the existence of weak solutions) can be found in the papers by 
Lions, Temam and Wang \cite{LTW1,LTW2,LTW3} and Lewandowski \cite{Lewan07}. Considering the technical 
difficulties of this 
new work, which has been solved on a cylindrical domain with Neumann-type boundary conditions on the 
top and bottom, these results must be classed as a major advance\footnote{\sf Unfortunately the methods 
used for HPE do not extend to the three-dimensional Navier-Stokes equations; although HPE had been 
thought to be the more difficult of the two. See Lions, Temam and Wang \cite{LTW1,LTW2,LTW3}.}. The 
significance of this result in establishing the existence and uniqueness of strong solutions of
viscous HPE is that \textbf{no singularities of any type can form in the solutions.} 

The title and abstract of this review refer to the idea of `extreme events'\,: this term should not 
be interpreted in the present context in a statistical sense, but as the spontaneous appearance and 
disappearance of local, sudden, intermittent events with large gradients in the atmosphere or the 
oceans. These events would register as spatially localized, intense concentrations of vorticity and
strain. Until the the Cao-Titi result, a major open question in the subject had been whether these 
strong variations ultimately remained smooth? Now that we know that solutions of HPE must remain smooth 
at every scale, the question still remains whether this smoothness precludes the existence of extreme 
events representing front-like structures\,: this would be manifest in jumps or steep 
gradients in double mixed derivatives such as $|\partial^{2}u_{1}/\partial x_{2}\partial x_{3}|$ or 
$|\partial^{2}u_{2}/\partial x_{1}\partial x_{3}|$ within \textit{finite} regions of space-time. 
The ultimate aim is to explain fine-structure processes such as those that occur in fronto-genesis\,:
see the work of Hoskins \cite{H75,Hrev82} and Hoskins and Bretherton \cite{HB72} for general background 
on this phenomenon.  

\S\ref{events} of this paper seeks to address this problem from a mathematical angle. The regularity of 
solutions opens a way of showing that front-like events are possible at very fine scales. The method 
used is to search for fine structure by examining the local behaviour of solutions in space-time. The 
conventional approach to PDEs has been to prove bounded-ness 
of norms which, as volume integrals, tend to obscure the fine structure of a solution.  The idea was 
first used (under strong regularity assumptions) to estimate intermittency in Navier-Stokes 
turbulence \cite{JDG09}. The mathematical approach is to show that solutions of HPE may be divided into 
two space-time regions $\mathbb{S}^{+}$ and $\mathbb{S}^{-}$. If $\mathbb{S}^{-}$ is non-empty then 
\textit{very large lower bounds} on double mixed derivatives of components of the velocity field 
$(u_{1},\,u_{2},\,u_{3})$ such as $|\partial^{2}u_{1}/\partial x_{2}\partial x_{3}|$ or 
$|\partial^{2}u_{2}/\partial x_{1}\partial x_{3}|$ may occur within $\mathbb{S}^{-}$. These large 
point-wise lower bounds on second-derivative quantities represent intense accumulation in the 
$(x_{1},x_{3})$- and $(x_{2},x_{3})$-planes, respectively.  Clearly, if solutions of great intensity 
were to accumulate in regions of the space-volume, this could only be allowed for finite times, in 
local spatial regions. Thus one would see the spontaneous formation of a front-like object localised 
in space that would only exist for a finite time.  More specifically, the lower bounds within 
$\mathbb{S}^{-}$ referred to above are a linear sum of $\mathcal{R}_{u_{hor}}^{6},~\mathcal{R}_{u_{3}}^{6}$ 
and $\mathcal{R}_{a,T}^{6}$ where $\mathcal{R}_{u_{hor}}$ and $\mathcal{R}_{u_{3}}$ are \textit{local} 
Reynolds numbers. That is, they are defined using the local space-time values of $u_{hor}(x_{1},x_{2},x_{3},
\tau) = \sqrt{u_{1}^{2} + u_{2}^{2}}$ and the vertical velocity $u_{3} = u_{3}(x_{1},x_{2},x_{3},\tau)$. 
Likewise, $\mathcal{R}_{a,T}$ is a local Rayleigh number, defined using the local temperature 
$T(x_{1},\,x_{2},\,x_{3},\,\tau)$. In contrast, $\Rey$ is the global Reynolds number. These lower bounds 
within the $\mathbb{S}^{-}$-regions can be converted into lower bounds on the two point-wise inverse 
length scales $\lambda_{H}$
\bel{lambdabd}
L\lambda_{H}^{-1} > c_{u_{hor}}\mathcal{R}_{u_{hor}} + c_{u_{3}}\mathcal{R}_{u_{3}} + c_{1,T}\mathcal{R}_{a,T}
+ c_{2,T}\Rey^{2/3}\mathcal{R}_{a,T}^{1/3} + \hbox{\small forcing}\,.
\ee
The term $\mathcal{R}_{u_{3}}\sim O(\varepsilon)$ is negligible compared to $\mathcal{R}_{u_{hor}}$ because
regions of strong vertical convection are absent in HPE. If such a result were valid for
NPE, it would be this vertical term that would be restored. The estimate for $\lambda_{H}$ in (\ref{lambdabd}) 
is of the order of a \textit{metre} or less, at high Reynolds numbers. Our interpretation of this very small 
length is not that it is necessarily  the thickness of a front, but that it may refer to the smallest scale 
of features \textit{within} a front. If an equivalent result for NPE could be found then the estimate of the 
length-scale $\lambda_{N}$ (say) would likely be considerably smaller than that for $\lambda_{H}$.

\section{\sf\textbf{Formulating the Primitive Equations}}\label{prim}

\begin{table}[htb]\sf
\bc
\begin{tabular}{||l|c|c||}\hline
\textit{Quantity} & \textit{Symbol} & \textit{Definition}\\\hline\hline
Typical horizontal length & $L$ &\\\hline 
Typical vertical length & $H$ &\\\hline
Typical temperature & $T_{0}$ &\\\hline
Aspect ratio & $\asp$ & $\asp = H/L$ \\\hline 
Typical velocity & $U_{0}$ & \\\hline
Twice vertical rotation rate & $f$ & \\\hline
Global Reynolds number & $\Rey$ & $\Rey = U_{0}L\nu^{-1}$\\\hline
Rossby number & $\varep$ & $\varep = U_{0}(Lf)^{-1}$\\\hline
Rayleigh number & $R_{a}$ & $R_{a} = g\alpha T_{0}H^{3}(\nu\kappa)^{-1}$\\\hline
Hydrostatic const & $a_{0}$ & $a_{0} = \varep \sigma\asp^{-3}R_{a}\Rey^{-2}$\\\hline
Frequency & $\omega_{0}$ & $\omega_{0} = U_{0}L^{-1}$\\\hline\hline
\end{tabular}
\caption{\small Definition of symbols for the primitive equations.}
\label{table1}
\ec
\end{table}
\begin{table}[htb]\sf
\bc
\begin{tabular}{||l|c|c|c||}\hline
 & \textit{Dimensionless} & \textit{Dimensional} & \textit{Relation}\\\hline\hline
Horiz co-ords & $x,y$ & $x_{1},x_{2}$ & $(x,\,y) = (x_{1},x_{2})L^{-1}$\\\hline
Vertical co-ord & $z$ & $x_{3}$ & $z = x_{3}H^{-1}$\\\hline
Time & $t$ & $\tau$ & $t = \tau U_{0}L^{-1}$\\\hline
Horiz velocities & $(u,\,v)$ & $\bu = (u_{1},\,u_{2})$ & $(u,\,v) = (u_{1},u_{2})U_{0}^{-1}$\\\hline
Vertical velocity & $w$ & $u_{3}$ & $\varep U_{0}\asp w = u_{3}$\\\hline
Velocity vector & $\bV = (u,\,v,\,\varep w)$ & &\\\hline
Hydro-Velocity & $\bv = (u,\,v,\,0)$& &\\\hline
Non-Hydro-Vel & $\bfv = (u,\,v,\,\asp^{2}\varep w)$& &\\\hline
Temperature & $\Theta$ & $T$ & $\Theta = T T_{0}^{-1}$\\\hline
$2D$-Gradient & $\nabla_{2} = \bivec \partial_{x} + \bjvec\partial_{y}$ & & \\\hline
$3D$-Gradient & $\nabla_{3} = \bivec \partial_{x} + \bjvec\partial_{y} + \bk\partial_{z}$ & 
$\nabla = \bivec \partial_{x_{1}} + \bjvec\partial_{x_{2}} + \bk\asp\partial_{x_{3}}$ & $\nabla_{3} = L\nabla$\\\hline
$3D$-Laplacian & $\Delta_{3} = \partial^{2}_{x} + \partial^{2}_{y} + \partial^{2}_{z}$ & & \\\hline
Vorticity & $\bom = \hbox{curl}\,\bV$ & $\bcapom$ & $\Omega_{3}
= \omega_{3}\omega_{0}$\\\hline 
Hydro-vorticity & $\bzeta = \hbox{curl}\,\bv$ &  &\\\hline 
Non-hydro-vorticity & $\bvarpi = \hbox{curl}\,\bfv$ &  &\\\hline 
Heat transport & $q$ & $Q$ & \\\hline
\hline
\end{tabular}
\caption{ Connection between dimensionless \& dimensional variables.}\label{table2}
\ec
\end{table}
In what follows, \textit{dimensionless} co-ordinates are denoted by $(x,\,y,\,z,\,t)$. These are 
related to \textit{dimensional} variables $(x_{1},x_{2},x_{3},\tau)$ through the horizontal length 
scale $L$, the vertical length scale $H$\,: Table 1 gives the various standard definitions.  
Table 2 gives the notation concerning dimensionless and dimensional variables and some of 
the relations between them. 
\par\smallskip
Dimensionless versions of both the HPE and NPE are expressed in terms of a set of velocity vectors 
based on the two horizontal velocities $(u,\,v)$ and the vertical velocity $w$.  
These form the basis for the three-dimensional vector%
\footnote{\sf Comparing with the notation in \cite{Holm96}, $(u,\,v) \equiv \bu_{2}$ while 
$\bV \equiv \bu_{3}$ and $\bfv \equiv \bv_{3}$.}
\bel{u3vdef}
\bV(x,y,z,t) = (u,\,v,\,\varep w)\,,
\ee
satisfying $\hbox{div}\,\bV = 0$ or $u_{x} + v_{y} = -\varep w_{z}$. The hydrostatic velocity vector 
is defined by
\bel{Hvel}
\bv = (u,\,v,\,0)\,,
\ee
and the non-hydrostatic velocity vector by
\bel{NHvel}
\bfv = (u,\,v,\,\asp^{2}\varep w)\,,
\ee
where $\asp$ is the aspect ratio defined in Table \ref{table3}. Neither $\bv$ nor $\bfv$ 
are divergence-free. 

The HPE that are studied in this paper should be compared with the modellers' HPE (MHPE) 
used by Richardson and incorporated in many numerical weather prediction and climate 
simulation models to this day \cite{DaCuMaMaStWhWo05,WhHoRoSt05,Cullen06,Cullen07}. 
While the MHPE are written for a compressible fluid in 
the non-Euclidean geometry of a shallow atmosphere (i.e. the domain of flow is a 
spherical shell of constant radius, although vertical extent and motion are allowed) 
the HPE as laid out in Section \ref{HPE-sec} are 
written for an incompressible fluid in a Cartesian geometry. The `incompressibility' of 
hydrostatic models appears when pressure is used as vertical coordinate. Hoskins and 
Bretherton \cite{HB72} used a function of pressure to define a vertical coordinate that 
delivers the hydrostatic primitive equations for a compressible fluid that are nearly 
isomorphic to those for an incompressible fluid when ordinary height is used as vertical 
coordinate. They then introduced an approximation in the continuity equation that made the near 
isomorphism exact. The resulting HPE have frequently been used in theoretical studies and 
have led to many important results, but they are not identical to the MHPE that numerical 
modellers since Richardson have typically used.

\subsection{\sf\textbf{Hydrostatic primitive equations (HPE)}}\label{HPE-sec}

The HPE in dimensionless form for the two horizontal velocities $u$ and $v$ are
\bel{PE1u}
\varep\left(\frac{\partial~}{\partial t} + \bV\cdot\nabla_{3}\right)u - v = 
\varep\Rey^{-1}\Delta_{3}u - p_{x}\,,
\ee
and
\bel{PE1v}
\varep\left(\frac{\partial~}{\partial t} + \bV\cdot\nabla_{3}\right)v + u = 
\varep\Rey^{-1}\Delta_{3}v - p_{y}\,.
\ee
There is no evolution equation for the vertical velocity $w$ which lies in both $\bV\cdot\nabla_{3}$ 
and the incompressibility condition $\hbox{div}\,\bV = 0$. The $z$-derivative of the pressure field 
$p$ and the dimensionless temperature $\Theta$ enter the problem through the hydrostatic equation
\bel{hydro1}
a_{0}\Theta + p_{z} = 0\,,
\ee
where $a_{0}$ is defined in Table 1. The hydrostatic velocity field $\bv = (u,\,v,\,0)$ appears 
when (\ref{PE1u}), (\ref{PE1v}) and (\ref{hydro1}) are combined
\bel{3DB}
\varep\left(\frac{\partial~}{\partial t} + \bV\cdot\nabla_{3}\right)\bv 
+ \bk\times\bv + a_{0}\bk\Theta = \varep\Rey^{-1}\Delta_{3}\,\bv - \nabla_{3} p\,,
\ee
and taken in tandem with the incompressibility condition $\hbox{div}\,\bv = -\varep w_{z}$. 
The dimensionless temperature%
\footnote{\sf Note that we use the upper-case $\Theta$ for the temperature to avoid confusion with 
lower-case $\theta$ which is conventionally used for the potential temperature.}
$\Theta$, with a specified heat transport term $q(x,y,z,t)$, 
satisfies
\bel{tempPDE}
\left(\frac{\partial~}{\partial t} + \bV\cdot\nabla_{3}\right)\Theta = 
(\sigma\Rey)^{-1}\Delta_{3}\Theta + q\,.
\ee
The chosen domain is a cylinder of height $H$ ($0 \leq z \leq H$) and radius $L$, 
designated as $C(L,H)$. In terms of dimensionless variables, boundary conditions are 
taken to be\,:
\ben
\item  On the top and bottom of the cylinder $z=0$ and $z=H$ the normal derivatives satisfy 
$u_{z} = v_{z} = \Theta_{z} = 0$ together with $w=0$. This is appropriate for the atmosphere 
where the heat transport term $q(x,\,y,\,t)$ in (\ref{tempPDE}) is dominant.  Ideally, in 
the case of the ocean, as in \cite{CT05,CT07}, the zero conditions for $u_{z}$ and $v_{z}$ 
on the cylinder top would need to be replaced by a specified wind stress, but this creates 
surface integrals which can only be estimated in terms of higher derivatives within the 
cylinder.

\item Periodic boundary conditions on all variables are taken on the side of the cylinder 
$S_{C}$, which is a minor change from those in \cite{CT05,CT07,NJ07}.
\een
The key feature for HPE is the use of enstrophy (i.e. vorticity squared) rather than kinetic energy as a 
quadratic measure of motion. This means that the pressure gradient is excluded from the dynamics at an early 
stage and, more importantly, that the analysis is immediately in terms of spatial derivatives of velocity 
components rather than the components 
themselves. In contrast, earlier approaches based available potential energy include the pressure field
until a global spatial integral removes it and this has made local results more difficult to reach.
In addition, energy arguments deliver results about attainable velocities not about velocity gradients.

The vorticity $\bvom = \mbox{curl}\,\bv$ is specifically given by
\bel{varpdef} 
\bvom = \hbox{curl}\,\bv = -\bivec v_{z} + \bjvec u_{z} + \bk(v_{x}-u_{y})\,.
\ee
This contains the same $\bk$-component as the full 3D-vorticity $\bom = \nabla_{3}\times\bV$ 
but with the $w$-terms missing. The relation between $\bzeta$ and $\bom$ gives rise to a 
technically important relation 
\bel{3DF1}
- \bV\cdot\nabla_{3}\bv = \bV\times\bvom -\shalf\nabla_{3}\big(u^{2} + v^{2}\big)\,.
\ee
Taking the curl of (\ref{3DB}) gives 
\beq{3DF2}
\varep\frac{\partial\bvom}{\partial t} &=& \varep\Rey^{-1}\Delta_{3}\bvom 
+ \varep\hbox{curl}\,\big(\bV\times\bvom) - \hbox{curl}\,(\bk\times\bv + a_{0}\bk\Theta)\,.
\eeq
This relation forms the basis of the proof of Theorem \ref{hydrothm} in \S\ref{events}.

\subsection{\sf\textbf{Non-hydrostatic primitive equations (NPE)}}

The NPE restore vertical acceleration so that, in contrast to (\ref{hydro1}), the equation 
for $w$ reads
\bel{wequn1}
\asp^{2}\varep^{2}\left(\frac{\partial~}{\partial t} + \bV\cdot\nabla_{3}\right)w 
+ a_{0}\Theta + p_{z} = \varep\Rey^{-1}\Delta_{3}w\,.
\ee
Using the definition of $\bfv$ in (\ref{NHvel}) and putting (\ref{wequn1}) together 
with (\ref{PE1u}) and (\ref{PE1v}) we find
\bel{nhPE}
\varep\left(\frac{\partial~}{\partial t} + \bV\cdot\nabla_{3}\right)\bfv 
+ \bk\times\bfv + a_{0}\bk\Theta = \varep\Rey^{-1}\Delta_{3}\,\bfv - \nabla_{3} p\,.
\ee
The equation for the temperature (\ref{tempPDE}) remains the same. In contrast to the 
hydrostatic case where regularity has been established with realistic Neumann-type 
boundary conditions on the top of a cylinder, several major open questions remain 
in the non-hydrostatic case\,:
\ben
\item No proof yet exists for the regularity of the NPE (\ref{nhPE}). Their prognostic 
equation for $w$ brings NPE much closer to the three-dimensional Navier-Stokes equations 
for a rotating stratified fluid than HPE, in which $w$ is diagnosed 
from the incompressibility condition. While regularity properties for Navier-Stokes 
fluids on very thin domains ($0< \asp <<1$) has been established \cite{thin1,thin2}, 
the general regularity problem remains open. 

\item The boundary integrals for HPE estimated on the top and bottom of the cylinder 
$C(H,L)$ are also problematic for NPE. Whereas the vorticity $\bzeta$ in the hydrostatic 
case has only terms $u_{z}$ and $v_{z}$ in the horizontal components, which are 
specified on the top and bottom of $C(H,L)$, the non-hydrostatic vorticity 
\bel{varpidef}
\bvarpi = \hbox{curl}\,\bfv
\ee 
also includes $w_{x}$ and $w_{y}$ terms which are not specified.
\een
Similar to (\ref{3DF1}), there exists a technical relation between $\bV$, $\bfv$ and 
$\bvarpi$
\bel{NH2}
- \bV\cdot\nabla_{3}\bfv = \bV\times\bvarpi -\shalf\nabla_{3}\big[u_{3}^{2} - 
\varep^{2}w^{2}(\asp^{2} - 1)^{2}\big]\,,
\ee
so the curl-operation on (\ref{nhPE}) gives 
\beq{NH3}
\varep\frac{\partial\bvarpi}{\partial t} &=& \varep\Rey^{-1}\Delta_{3}\bvarpi
+ \varep\hbox{curl}\,\big(\bV\times\bvarpi) - \hbox{curl}\,(\bk\times\bfv + a_{0}\bk\Theta)
\eeq
without involving pressure terms.
\par\medskip
To address the influence of the material derivative in (\ref{wequn1}), typical 
values of $\asp^{2}\varep^{2}$ can be determined. Typical observed values for 
mid-latitude synoptic weather and climate systems are: 
\beq{typical}
\asp &=& H/L \approx 10^4 m/10^6 m \approx 10^{-2}\non\\
W/U &\approx& 10^{-2} ms^{-1}/10 ms^{-1} \approx 10^{-3}\\
\varep &=& U/(f_0L) \approx 10 ms^{-1}/(10^{-4} s^{-1}10^6 m) \approx 10^{-1}\,.\non 
\eeq
For mid-latitude large-scale ocean circulation, the corresponding numbers are:
\beq{H2}
\asp &=& H/L \approx 10^3 m/10^5 m \approx 10^{-2}\,\non\\
W/U &\approx& 10^{-3} ms^{-1}/10^{-1} ms^{-1} \approx 10^{-2}\\
\varep &=& U/(f_0L) \approx 10^{-1} ms^{-1}/(10^{-4} s^{-1}10^5 m) \approx 10^{-2}\,.\non
\eeq
Thus, for these typical conditions, $\asp^{2}\varep^{2}\approx 10^{-8}-10^{-6}\ll1$ 
so the hydrostatic approximation can be expected to be extremely accurate in calculations 
of either synopic weather and climate, or large-scale ocean circulation at mid-latitude\,: 
see \cite{RH1,RH2,RH3}. 

\section{\sf\textbf{Extreme events \& their interpretation}}\label{events}

The proof of global existence and uniqueness of solutions of HPE by Cao and Titi guarantees 
that at each point in space time a unique solution exists \cite{CT05,CT07,NJ07}. 
This result can be exploited to prove Theorem 1 and from this to consider the idea of 
extreme events.
\begin{table}[htb]\sf
\bc
\begin{tabular}{||l|c||}\hline
\textit{HPE quantity} & \textit{Definition} \\\hline\hline
Local (horizontal) Reynolds no in terms of $|\bu|$ & $\mathcal{R}_{u_{hor}} = L |\bu|\nu^{-1}$\\\hline
Local (verterical) Reynolds no in terms of $|u_{3}|$ & $\mathcal{R}_{u_{3}} = L |u_{3}|\nu^{-1}$\\\hline
Local Rayleigh no in terms of $T$ & $\mathcal{R}_{a,T} = g\alpha H^{3}T (\nu\kappa)^{-1}$\\\hline
$\beta_{1}$ & $\beta_{1} = 2L^{2}\delta_{3}\omega_{0}^{-2}$\\\hline
$\beta_{2}$ & $\beta_{2} = L^{4}(1-2\delta_{3})\sigma^{-1} T_{0}^{-2}$\\\hline
$\beta_{u_{hor}}$ & $\beta_{u_{hor}} = \frac{9}{4\delta_{1}^{3}} + \frac{243}{32\delta_{3}^{3}}$\\\hline
$\beta_{u_{3}}$ & $\beta_{u_{3}} = \left(\frac{3}{2\delta_{1}^{3}} 
+ \frac{\sigma^{2}}{12\delta_{3}^{3}}\right)$\\\hline
$\beta_{1,T}$ & $\beta_{1,T} = \frac{\sigma^2}{6}\left(1 
+ \frac{\sigma}{64}\right)\asp^{-6}\delta_{3}^{-3}$\\\hline
$\beta_{2,T}$ & $\beta_{2,T} = \frac{\varep^{-2}\asp^{-6}}{2\delta_{2}}$\\\hline
$\mathcal{V}_{*}$ & $\mathcal{V}_{*} = \pi L^{2} H \tau_{*}$\\\hline 
\end{tabular}
\par\vspace{0mm}
\caption{\small Definitions in the hydrostatic case.}\label{table3}
\ec
\end{table}
\par\smallskip\noindent
Consider the arbitrary positive constants $0 < \delta_{1},~\delta_{2} < 1$ chosen such that
\bel{del3def}
2\delta_{3} = 1- \threequart\delta_{1}-\delta_{2} > 0\,. 
\ee
These are used to define a hydrostatic forcing function $F_{H}(Q)$\,: 
\bel{Fdef} 
F_{H}(Q) = 
\varep^{-2}\left(\delta_{2} + \frac{1}{2\delta_{2}}\right)\Rey^{2}\mathcal{R}_{u_{hor}}^{2} 
+ \frac{\sigma}{6\delta_{3}T_{0}^{2}\omega_{0}^{2}}\Rey |Q|^{2}\,.
\ee
The spatially global quantity $\mathcal{H}(t)$ 
\bel{Hdef}
\mathcal{H}(t) = \I\left(\Rey^{3} |\bvom|^{2} + \Rey |\Theta_{z}|^{2}\right)dV\,.
\ee 
is now used in the proof of the following theorem\,:
\begin{theorem}\label{hydrothm} In $C(H,L)$ over some chosen time interval $[0,\,\tau_{*}]$, 
the space-time 4-integral satisfies
\beq{hthmequn}
\int_{0}^{\tau_{*}}\I \Big\{\!\!\!\!&-\beta_{1}&\!\!\!\!
\left[ |\nabla_{2}\Omega_{3}|^{2} 
+ \asp^{2}\left|\frac{\partial^{2}u_{1}}{\partial x_{2}\partial x_{3}}\right|^{2} 
+ \asp^{2}\left|\frac{\partial^{2}u_{2}}{\partial x_{1}\partial x_{3}}\right|^{2}\right]
- \beta_{2}\asp^{2}|\nabla_{3}T_{x_{3}}|^{2}
+ \beta_{u_{1}}\mathcal{R}_{u_{hor}}^{6}\\ &+& \beta_{u_{3}}\mathcal{R}_{u_{3}}^{6} 
+ \beta_{1,T}\mathcal{R}_{a,T}^{6} + \beta_{2,T}\Rey^{4}\mathcal{R}_{a,T}^{2} + 
F_{H}(Q) + \shalf \mathcal{V}_{*}^{-1}\mathcal{H}(0)\Big\}\,dx_{1}dx_{2}dx_{3} d\tau > 0\non
\eeq
where the coefficients are given in Table \ref{table3}.
\end{theorem}
\textbf{Remark 1\,:}  The proof in \S \ref{hydroproof} shows that the right hand side of (\ref{hthmequn})
is, in fact, $\shalf\mathcal{H}(\tau_{*})$. Because solutions of HPE are regular this is bounded above
for all values of $\tau_{*}$. However, it has a lower bound of zero which, although not necessarily a 
good lower bound, is uniform in $\tau_{*}$. Regularity of solutions is also a necessity in order to 
extract point-wise functions from within the 4-integral in (\ref{hthmequn}).
\par\medskip\noindent
\textbf{Remark 2\,:} Of course, any integral of the form $\int\!\!\int (A-B)\,dVdt > 0$ trivially 
indicates that there must be regions of space-time where $A > B$ but, potentially, there could 
also be regions where $A \leq B$. Such integrals are common, particularly energy integrals, and 
in most cases the information gained is of little interest. In the case of (\ref{hthmequn}), 
however, the proof in \ref{hydroproof} will demonstrate that much effort has gone into rigorously 
manipulating it into a form that produces sensible and recognizable physics. From (\ref{hthmequn}) 
we conclude that\,: 
\ben
\item There are regions of space-time $\mathbb{S}^{+}\subset \mathbb{R}^{4}$ on which 
\beq{plus}
&\beta_{1}&\!\!\!\left[|\nabla_{2}\Omega_{3}|^{2} 
+ \asp^{2}\left|\frac{\partial^{2}u_{1}}{\partial x_{2}\partial x_{3}}\right|^{2} 
+ \asp^{2}\left|\frac{\partial^{2}u_{2}}{\partial x_{1}\partial x_{3}}\right|^{2}\right]
+ \beta_{2}\asp^{2}|\nabla T_{x_{3}}|^{2}\non\\ 
& < & \beta_{u_{hor}}\mathcal{R}_{u_{hor}}^{6} + \beta_{u_{3}}\mathcal{R}_{u_{3}}^{6} 
+ \beta_{1,T}\mathcal{R}_{a,T}^{6} 
+\beta_{2,T}\Rey^{4}\mathcal{R}_{a,T}^{2} + F_{H}(Q) + O(\tau_{*}^{-1})
\eeq

\item Potentially  there are also regions of space-time $\mathbb{S}^{-}\subset \mathbb{R}^{4}$ 
on which
\beq{minus}
&\beta_{1}&\!\!\!\left[|\nabla_{2}\Omega_{3}|^{2} 
+ \asp^{2}\left|\frac{\partial^{2}u_{1}}{\partial x_{2}\partial x_{3}}\right|^{2} 
+ \asp^{2}\left|\frac{\partial^{2}u_{2}}{\partial x_{1}\partial x_{3}}\right|^{2}\right]
+ \beta_{2}\asp^{2}|\nabla T_{x_{3}}|^{2}\non\\ 
&\geq & \beta_{u_{hor}}\mathcal{R}_{u_{hor}}^{6} + 
\beta_{u_{3}}\mathcal{R}_{u_{3}}^{6} + \beta_{1,T}\mathcal{R}_{a,T}^{6} 
+ \beta_{2,T}\Rey^{4}\mathcal{R}_{a,T}^{2} + F_{H}(Q) + O(\tau_{*}^{-1})
\eeq
\een
It must be stressed that there is no information here about the nature of the two sets 
$\mathbb{S}^{\pm}$\,: the 4-integral in (\ref{hthmequn}) above yields no more information 
other than the possibility of a non-empty set $\mathbb{S}^{-}$ existing. It says nothing 
about the spatial or temporal statistics of the subsets of $\mathbb{S}^{-}$ (which may 
have a very sensitive $\tau_{*}$-dependence) nor does it give any indication of their topology. 
\textbf{These results, however, are consistent with the observations of fronts in the 
atmosphere where large second gradients appear spontaneously in confined spatial regions 
often disappearing again in an equally spontaneous manner.} This behaviour would also be 
consistent with $\mathbb{S}^{-}$ being comprised of a disjoint union of subsets although 
no details can be deduced from (\ref{hthmequn}).

To illustrate the nature of a front, very large values of double mixed-derivatives are required in 
local parts of the flow as envisaged in the early and pioneering work of Hoskins \cite{H75,Hrev82}. 
For instance, very large values of $|\partial^{2}u_{1}/\partial x_{3}\partial x_{2}|^{2}$ or 
$|\partial^{2}u_{2}/\partial x_{3}\partial x_{1}|^{2}$ would represent intense accumulation in the 
$(x_{1},x_{3})$- and $(x_{2},x_{3})$-planes respectively. 
\par\smallskip
The large lower bounds in Theorems \ref{hydrothm} can be interpreted in terms of a length scale. 
To achieve this, define the point-wise inverse length scale 
$\lambda_{H}^{-1}$ such that%
\footnote{\sf The point-wise local length scale $\lambda_{H}^{-1}$ is formed in the same 
dimensional manner as the Kraichnan length $\ell_{k}$, namely, from a combination of the 
palenstrophy $|\nabla\bom|^{2}$and the viscosity $\nu$ given by $\ell_{k}^{-6} = 
\nu^{-2}|\nabla\bom|^{2}$.}
\bel{lsH}
\big(L\lambda_{H}^{-1}\big)^{6}
= L^{2}\omega_{0}^{-2}
\left(|\nabla_{2}\Omega_{3}|^{2} + \asp^{2}\left|\frac{\partial^{2}u_{1}}{\partial x_{2}\partial x_{3}}\right|^{2} 
+ \asp^{2}\left|\frac{\partial^{2}u_{2}}{\partial x_{1}\partial x_{3}}\right|^{2}\right) 
+ \asp^{2}\beta_{2}\beta_{1}^{-1}L^{4}|\nabla_{3}T_{x_{3}}|^{2}T_{0}^{-2}\,,
\ee
The sixth-powers on the right hand sides of Theorem \ref{hydrothm} gives results within $\mathbb{S}^{-}$
\bel{lbHN}
L\lambda_{H}^{-1} > c_{u_{hor}}\mathcal{R}_{u_{hor}} + c_{u_{3}}\mathcal{R}_{u_{3}} + c_{1,T}\mathcal{R}_{a,T} 
+ c_{2,T}\Rey^{2/3}\mathcal{R}_{a,T}^{1/3} + \hbox{forcing}
\ee
where the coefficients $c_{u_{hor}},~c_{u_{3}}$ and $c_{i,T}$ can be calculated from Theorems 
\ref{hydrothm}\,: indeed the term $\mathcal{R}_{u_{3}}$ can be ignored as the vertical velocity 
$u_{3} \sim O(\varepsilon)$. (\ref{lbHN}) can be interpreted as a lower bound on the inverse 
length scale of the smallest feature in a front. As already noted, no information is available 
on any statistics nor on the shape and size of subsets of $\mathbb{S}^{-}$.

\subsection{\sf\textbf{Proof of Theorem \ref{hydrothm}~}}\label{hydroproof}

\subsubsection{\sf The evolution of the enstrophy}\label{pv}

The equation for the HPE enstrophy $\bvom$ in (\ref{3DF2}) is now considered on the cylinder $C(H,L)$ 
with the boundary conditions $u_{z} = v_{z} = w = 0$ on the top and bottom but with periodic 
side-wall conditions
\beq{3DG}
\shalf\varep\frac{d~}{dt}\I |\bvom |^{2}\,dV &=& \varep\Rey^{-1}\I\bvom\cdot\Delta_{3}\bvom\,dV 
+ \varep \I\bvom\cdot \hbox{curl}\,\big(\bV\times\bvom\big)dV\non\\ 
&-& \I \bvom\cdot\hbox{curl}(\bk\times\bv + a_{0}\bk\Theta)\,dV\,.
\eeq
The notation in the rest of the paper is
\bel{not1}
\|f\|_{p} = \left(\I |f|^{p}\,dV\right)^{1/p}\,.
\ee
Note that in estimating the integrals in (\ref{3DG}), a standard vector identity 
and the Divergence Theorem are used in which a surface integral naturally appears 
each time. However, on the cylinder top and bottom
\bel{BC1}
\bk\times\bvom = 0\qquad\mbox{on}\qquad z = 0,~H\,.
\ee
This and the side-wall periodic boundary conditions make zero all the surface integrals 
that appear from the Divergence Theorem.

Considering (\ref{3DG} term by term, the first is ($\mbox{div}\,\bvom = 0$)
\bel{3DH}
\I\bvom\cdot\Delta_{3}\bvom\,dV = -\I\bvom\cdot\hbox{curl\,curl}\,\bvom\,dV 
= -\I |\hbox{curl}\,\bvom|^{2}\,dV\,. 
\ee
The second integral on the right hand side of (\ref{3DG}) 
\beq{mat1} 
\I\bvom\cdot\hbox{curl}\,(\bV\times\bvom)dV = 
\I\hbox{curl}\,\bvom\cdot(\bV\times\bvom)dV\,,
\eeq
and so 
\beq{mat2}
\left|\I\bvom\cdot\hbox{curl}\,(\bV\times\bvom)\,dV\right| 
&\leq& \|\hbox{curl}\,\bvom\|_{2}\|\bV\|_{6}\|\bvom\|_{3}\,.
\eeq
Thirdly, 
\beq{3DJ}
\I \bvom\cdot\hbox{curl}(\bk\times\bv + a_{0}\bk\Theta)\,dV &=& 
\I \hbox{curl}\,\bvom\cdot(\bk\times\bv + a_{0}\bk\Theta)\,dV\,.
\eeq
Thus (\ref{3DG}) can be re-written as
\beq{3DK}
\shalf\varep\frac{d~}{dt}\I |\bvom|^{2}\,dV &\leq& -\varep\Rey^{-1}\I|\hbox{curl}\bvom |^{2}\,dV 
+ \varep\|\hbox{curl}\,\bvom\|_{2}\|\bV\|_{6}\|\bvom\|_{3}\non\\
&-& \I(\hbox{curl}\,\bvom)\cdot\left(\bk\times\bv + a_{0}\bk\Theta\right)\,dV\non\\
&\equiv& -\varep\Rey^{-1}\I |\hbox{curl}\bvom |^{2}\,dV + \hbox{T1 + T2 + T3}\,,
\eeq
where
\bel{T12}
\hbox{T1} = \varep\|\hbox{curl}\,\bvom\|_{2}\|\bV\|_{6}\|\bvom\|_{3}
\ee
\bel{T34}
\hbox{T2} = -\I(\hbox{curl}\bvom)\cdot(\bk\times\bv)\,dV\,,
\qquad\qquad
\hbox{T3} = - a_{0}\I(\hbox{curl}\bvom)\cdot (\bk\Theta)\,dV\,. 
\ee
Before estimating T1, T2, and T3, we prove\:
\par\smallskip
\begin{lemma}\label{zetalem}
In the cylinder $C(H,L)$
\bel{T1I}
\|\bvom\|_{3} \leq 3^{1/2}\|\hbox{curl}\,\bvom\|_{2}^{1/2}\|\bv\|_{6}^{1/2}\,.
\ee
\end{lemma}
\textbf{Proof\,:} Consider
\beq{T1F}
\I |\bzeta|^{3}dV &=& \I \bvom\cdot(\zeta\bvom)dV\non\\ 
&=& \I\bv\cdot\hbox{curl}\,(\zeta\bvom)dV - \int_{S}\hbox{div}\,
\big(\zeta(\bv\times\bzeta)\big)dS\non\\ 
&=&  \I\bv\cdot(\zeta\hbox{curl}\,\bvom + (\nabla_{3}\zeta)\times\bvom)dV
 - \int_{S}\hbox{div}\,\big(\zeta(\bv\times\bzeta)\big)dS\,.
\eeq
The surface integral is zero because $\bk\times \bvom = 0$ on 
the cylinder top and bottom. A vector identity
\bel{T1Ga}
\nabla_{3}\zeta = \shalf(\zeta)^{-1}\nabla_{3}(\bvom\cdot\bvom) = 
\hat{\bvom}\cdot\nabla_{3}\bvom + \hat{\bvom}\times\hbox{curl}\,\bvom
\ee
allows us to write
\beq{T1Gb}
\hbox{curl}\,(\zeta\bvom) 
= \zeta\hbox{curl}\,\bzeta + (\bvom\times\hbox{curl}\,\bvom)\times\hat{\bvom} - 
\bvom\times(\hat{\bvom}\cdot\nabla_{3}\bvom)\,,
\eeq
so we conclude that
\bel{T1Ha}
\I |\bzeta|^{3}dV = \I \bv\cdot\left(\zeta\hbox{curl}\,\bzeta + 
(\bvom\times\hbox{curl}\,\bvom)\times\hat{\bvom} - 
\bvom\times(\hat{\bvom}\cdot\nabla_{3}\bvom)\right)\,dV\,.
\ee
Using a H\"older inequality, it is then found that 
\bel{T1Hb}
\I |\bzeta|^{3}dV \leq 3\|\hbox{curl}\,\bvom\|_{2}\|\bvom\|_{3}\|\bv\|_{6}
\ee
giving the result (\ref{T1I}). \hfil $\square$
\par\medskip
\begin{lemma}\label{Tlem}
Within the cylinder $C(L,H)$, T1, T2 and T3 are estimated as
\beq{T1est}
|\hbox{T1}\,| &\leq& \threequart\delta_{1}\varep \Rey^{-1}\|\hbox{curl}\,\bvom\|_{2}^{2} 
+ \frac{3\varep}{4\delta_{1}^{3}}\Rey^{3}\left(2\|\bV\|_{6}^{6} + \|\bv\|_{6}^{6}\right)
\eeq
for any $\delta_{1} > 0$. Moreover, for any $\delta_{2} > 0$ and $\delta_{2} > 0$ 
\bel{T2est}
|\hbox{T2}\,|\leq \shalf\delta_{2}\varep\Rey^{-1}\|\hbox{curl}\,\bvom\|_{2}^{2} 
+ \shalf\delta_{2}^{-1}\varep^{-1}\Rey\|\bv\|_{2}^{2}\,,
\ee
\bel{T3est}
|\hbox{T3}\,| \leq \shalf\delta_{2}\varep\Rey^{-1}\|\hbox{curl}\,\bvom\|_{2}^{2} 
+ \frac{a_{0}^2}{2\varep\delta_{2}}\Rey\|\Theta\|_{2}^{2}\,.
\ee
\end{lemma}
\textbf{Proof\,:} In the following the $\delta_{i} > 0$ are constants introduced by a series of 
Young's inequalities\,:
\par\medskip\noindent
1) Using Lemma \ref{zetalem} T1 can be written as 
\beq{TL1}
\hbox{T1} &=& \varep\|\hbox{curl}\,\bvom\|_{2}\|\bV\|_{6}\|\bvom\|_{3}\\
&\leq& \varep\|\hbox{curl}\,\bvom\|_{2}\|\bV\|_{6}
\times 3^{1/2}\|\hbox{curl}\,\bvom\|_{2}^{1/2}\|\bv\|_{6}^{1/2}\non\\
&\leq& \left(\delta_{1}\varep \Rey^{-1}\|\hbox{curl}\,\bvom\|_{2}^{2}\right)^{3/4}
\left(9\varep\delta_{1}^{-3}\Rey^{3}\|\bV\|_{6}^{6}\right)^{1/6}
\left(9\varep\delta_{1}^{-3}\Rey^{3}\|\bv\|_{6}^{6}\right)^{1/12}\non\\
&\leq& \threequart\delta_{1}\varep \Rey^{-1}\|\hbox{curl}\,\bvom\|_{2}^{2} 
+ \frac{3\varep}{4\delta_{1}^{3}}\Rey^{3}
\left(2\|\bV\|_{6}^{6} + \|\bv\|_{6}^{6}\right)\,.
\eeq
%
\par\medskip\noindent
2) From (\ref{T34}), T2 can be estimated as 
\beq{T3Ab}
|\hbox{T2}| &\leq& \|\hbox{curl}\,\bvom\|_{2}\|\bv\|_{2} = 
\left(\delta_{2}\varep\Rey^{-1}\|\hbox{curl}\,\bvom\|_{2}^{2}\right)^{1/2}
\left(\delta_{2}^{-1}\varep^{-1}\Rey\|\bv\|_{2}^{2}\right)^{1/2}\non\\
&\leq& \shalf\delta_{2}\varep\Rey^{-1}\|\hbox{curl}\,\bvom\|_{2}^{2} 
+ \shalf\delta_{2}^{-1}\varep^{-1}\Rey\|\bv\|_{2}^{2}\,.
\eeq
\par\medskip\noindent
3) From (\ref{T34}), and using the same constant $\delta_{2}$, T3 is estimated as 
\beq{T4Ab}
|\hbox{T3}| &\leq& a_{0}\|\hbox{curl}\,\bvom\|_{2}\|\Theta\|_{2} = 
\left(\delta_{2}\varep\Rey^{-1}\|\hbox{curl}\,\bvom\|_{2}^{2}\right)^{1/2}
\left(\frac{a_{0}^{2}}{\delta_{2}\varep}\Rey\|\Theta\|_{2}^{2}\right)^{1/2}\non\\
&\leq& \shalf\delta_{2}\varep\Rey^{-1}\|\hbox{curl}\,\bvom\|_{2}^{2} 
+ \frac{a_{0}^2}{2\delta_{2}\varep}\Rey\|\Theta\|_{2}^{2}\,.
\eeq 
as advertized. \hfil $\square$
\par\medskip\noindent
Returning to (\ref{3DG}), a division by $\varep$ and a gathering terms gives
\beq{comb2}
\shalf\frac{d~}{dt}\I |\bvom|^{2}dV
&\leq&  -\left( 1 - \threequart\delta_{1} - \delta_{2}\right)
\Rey^{-1}\I|\hbox{curl}\,\bvom|^{2}dV \\
&+& \frac{3}{4\delta_{1}^{3}}\Rey^{3}\left(2\varep^{6}\|w\|_{6}^{6} + 3\|\bv\|_{6}^{6}\right) 
+ \frac{1}{\varep^{2}}\left(\delta_{2} + \frac{1}{2\delta_{2}}\right)\Rey\|\bv\|_{2}^{2} 
+ \frac{a_{0}^2}{2\varep^{2}\delta_{2}}\Rey\|\Theta\|_{2}^{2}\,.\non
\eeq
\rem{
Thus we fix the $\delta_{i}$ such that $\delta_{3} = 
(1 - \frac{3}{4}\delta_{1} -\frac{1}{2}\delta_{2} - \delta_{2})$. In fact we choose 
\bel{delchoose}
\delta_{1} = \delta_{2} = \delta_{2} = 2/9\qquad\qquad\delta_{3} = 1/2
\ee 
to turn (\ref{comb2}) into
\beq{comb3}
\shalf\varep\frac{d~}{dt}\I |\bvom|^{2}dV &\leq& -\shalf \varep\Rey^{-1}\I \sum_{i=1}^{6} M_{i}(\bv)^{2}dV 
+ \frac{9^{3}}{96}\left(36\varep\Rey^{3} + \varep^{\alpha_{1}}\Rey^{a_{1}}\right)\|\bv\|_{6}^{6}\non\\ 
&+&  \frac{9^{3}}{48}\varep^{\alpha_{2}}\Rey^{a_{2}}\|w\|_{6}^{6}
+ \frac{9}{4\varep}\Rey\,\|\bv\|_{2}^{2} + \frac{9a_{0}^2}{4\varep}\Rey\,\|T\|_{2}^{2}\,,
\eeq
where $\alpha_{1} + 2\alpha_{2} = 15$ and $a_{1} + 2a_{2} = 9$ and the $M_{i}(\bv)$ are double mixed-derivatives
of $u$ and $v$ -- these are defined earlier in the document.  Note that we can choose the $a_{i}$ and 
$\alpha_{i}$ according to their constraints to weaken or strengthen terms as we wish. Note also that unhappy 
terms like $\|\bv_{zz}\|_{2}^{2}$ have vanished having been carefully cancelled by judicious choices of 
coefficients, which is the origin of the strange fractions. In fact, the two coefficients are $\frac{9^{3}}{96} 
\approx 7.5$ and $\frac{9^{3}}{48} \approx 15$. 
}
\par\smallskip\noindent
The following Lemma relates $\I|\hbox{curl}\,\bvom|^{2}dV$ to sums of squares.
\begin{lemma}\label{lemma1} Let $\omega_{3}$ be the third component of the full 
vorticity $\bom = \nabla_{3}\times\bV = (\omega_{1},\,\omega_{2},\,\omega_{3})$. 
Then for any $0 < \delta_{0} < 1$
\beq{leminequal}
\I |\hbox{curl}\,\bvom|^{2}dV &>& \min\{1,\,2(1-\delta_{0})\}\I|\nabla_{2}\omega_{3}|^{2}dV 
+ 2\delta_{0}\I \big\{|u_{yz}|^{2} + |v_{xz}|^{2}\big\}dV\non\\
&+& \I \big\{|u_{zz}|^{2} + |v_{zz}|^{2} + \varep^{2}(1-\delta_{0})|w_{zz}|^{2}\big\}dV\,.
\eeq
\end{lemma}
\par\medskip\noindent
\textbf{Proof\,:} Using $\bvom$ in (\ref{varpdef}) in the form $\bvom = -v_{z}\bivec 
+ u_{z}\bjvec + \omega_{3}\bk$ with $\omega_{3} = v_{x}-u_{y}$ and recalling 
that $\varep w_{z} = -(u_{x} + v_{y})$
\bel{lempr1}
\hbox{curl}\,\bvom = \left|
\begin{array}{rcc}
\bivec & \bjvec & \bk\\
\partial_{x} & \partial_{y} & \partial_{z}\\
-v_{z} & u_{z} & \omega_{3}
\end{array}\right| 
= \bivec(\omega_{3,y} -u_{zz}) - \bjvec(\omega_{3,x} + v_{zz}) -\varep \bk w_{zz}
\ee
Thus
\beq{lempr2}
|\hbox{curl}\,\bvom|^{2} = |\nabla_{2}\omega_{3}|^{2} + \big(u_{zz}^{2} + v_{zz}^{2} + 
\varep^{2}w_{zz}^{2}\big) -2\big(\omega_{3,y}u_{zz} - \omega_{3,x}v_{zz}\big)\,.
\eeq
Now, invoking the boundary conditions and the fact that $\omega_{3} = v_{x}-u_{y}$, 
integration by parts gives
\beq{lempr3}
-2 \I (\omega_{3,y}u_{zz} - \omega_{3,x}v_{zz}\big) dV = 2\I |\omega_{3,z}|^{2}dV 
\eeq
and so
\bel{lempr6}
\I|\hbox{curl}\,\bvom|^{2}dV = \I \big\{|\nabla_{2}\omega_{3}|^{2} + 2|\omega_{3,z}|^{2}dV\big\} 
+ \I\big\{u_{zz}^{2} + v_{zz}^{2} + \varep^{2} w_{zz}^{2}\big\}dV\,. 
\ee
Moreover, it is easily shown that
\beq{lempr7}
\I \big(\varep^{2}|w_{zz}|^{2} + 2|\omega_{3,z}|^{2} \big)\,dV 
&=& \I \big\{(u_{xz}-v_{yz})^{2} + 2(u_{yz}^{2}+v_{xz}^{2})\big\}\,dV\non\\
&>& 2\I (u_{yz}^{2}+v_{xz}^{2})\,dV
\eeq
where a pair of horizontal integrations by parts in the first line of (\ref{lempr7}) 
have been performed. A linear combination of (\ref{lempr6}) and (\ref{lempr7}) gives
\beq{lempr8}
\I|\hbox{curl}\,\bvom|^{2}dV &=& \I \big\{|\nabla_{2}\omega_{3}|^{2} + 2|\omega_{3,z}|^{2}dV\big\}
+ \I\big\{u_{zz}^{2} + v_{zz}^{2} + \varep^{2}w_{zz}^{2}\big\}dV\non\\
&>& \I \big\{|\nabla_{2}\omega_{3}|^{2} + 2(1-\delta_{0})|\omega_{3,z}|^{2}dV\big\}\non\\ 
&+& \I\big\{u_{zz}^{2} + v_{zz}^{2} + (1-\delta_{0})\varep^{2}w_{zz}^{2}\big\}dV + 
2\delta_{0}\I (u_{yz}^{2}+v_{xz}^{2})\,dV\,,
\eeq
which gives (\ref{leminequal}). This completes the proof. \hfil $\square$ 


\subsubsection{\sf The evolution of $\I|\Theta_{z}|^{2}dV$}\label{gradtemp}

The partial differential equation for $\Theta$ given in (\ref{tempPDE}) with BCs applied on $C(L,H)$ 
\bel{newtemp1}
\frac{\partial \Theta}{\partial t} + \bV\cdot\nabla_{3} \Theta = (\sigma\Rey)^{-1}\Delta_{3} \Theta + q
\ee
is now differentiated with respect to $z$ to give 
\beq{newtemp2}
\shalf\frac{d~}{dt}\I |\Theta_{z}|^{2}dV &=& (\sigma\Rey)^{-1}\I \Theta_{z}(\Delta_{3}\Theta_{z})dV
-\I \Theta_{z}\frac{\partial~}{\partial z}\left(\bV\cdot\nabla_{3}\Theta\right) dV + \I \Theta_{z}q_{z}dV\non\\
&=& -(\sigma\Rey)^{-1}\I |\nabla_{3} \Theta_{z}|^{2}dV 
- \I \bV\cdot\nabla_{3} (\shalf \Theta_{z}^{2})dV\non\\
&-& \I \Theta_{z}(u_{z}\Theta_{x} + v_{z}\Theta_{y} + \varep w_{z}\Theta_{z})dV + \I \Theta_{z}q_{z}dV\,.
\eeq
However, given that $\hbox{div}\,\bV = 0$ and $w = 0$ on $S^{\pm}$
\beq{newtemp3}
\I \bV\cdot\nabla_{3} (\shalf \Theta_{z}^{2})dV &=& \I \left\{\hbox{div}\,(\shalf \Theta_{z}^{2}\,\bV) 
-\shalf \Theta_{z}^{2}\,\hbox{div}\,\bV\right\}dV\non\\
&=& \shalf \int_{S}(\bn\cdot\bV) \Theta_{z}^{2}\,dS\non\\ 
&=& \pm \shalf \varep\int_{S^\pm}w \Theta_{z}^{2}\,dxdy = 0\,.
\eeq
Integrating by parts the 3rd and 4th terms in (\ref{newtemp2}) gives
\beq{newtemp4}
\shalf\frac{d~}{dt}\I |\Theta_{z}|^{2}dV 
&=& -(\sigma\Rey)^{-1}\I |\nabla_{3}\Theta_{z}|^{2}dV+\I\Theta_{z}(u_{xz}+v_{yz}+\varep w_{zz})\Theta
\,dV\non\\
&+& \I \Theta(u_{z}\Theta_{xz} + v_{z}\Theta_{yz} + \varep w_{z}\Theta_{zz})dV - \I \Theta_{zz} q dV\non\\
&=& -(\sigma\Rey)^{-1}\I |\nabla_{3} \Theta_{z}|^{2}dV + \I\Theta\left(u_{z}\Theta_{xz} 
+ v_{z}\Theta_{yz} + \varep w_{z}\Theta_{zz}\right)dV\non\\
&-& \I \Theta_{zz} q dV
\eeq
where $\hbox{div}\,\bV=0$ has been used. Using a H\"older inequality it is found that 
\beq{newtemp5}
\shalf\frac{d~}{dt}\I |\Theta_{z}|^{2}dV &\leq& -(\sigma\Rey)^{-1}\I |\nabla_{3} \Theta_{z}|^{2}dV + 
\|\Theta_{zz}\|_{2}\|q\|_{2}\non\\ 
&+&  \|\Theta\|_{6}\left\{\|u_{z}\|_{3}\|\Theta_{xz}\|_{2} + \|v_{z}\|_{3}\|\Theta_{yz}\|_{2} 
+ \varep\|w_{z}\|_{3}\|\Theta_{zz}\|_{2}\right\}\,.
\eeq
The next task, addressed in the following Lemma, is to estimate $\|u_{z}\|_{3}$, $\|v_{z}\|_{3}$ 
and $\|w_{z}\|_{3}$ in terms of their second derivatives. 
\begin{lemma}\label{L3lem} With Neumann boundary conditions on $C(H,L)$, 
the vector $\bv_{z}$ and the scalar $w_{z}$ satisfy
\bel{AppA1}
\|\bv_{z}\|_{3} \leq 6^{1/2}\|\bv_{zz}\|_{2}^{1/2}\|\bv\|_{6}^{1/2}\,,
\ee
\bel{AppA2}
\|w_{z}\|_{3} \leq 2^{1/2}\|w_{zz}\|_{2}^{1/2}\|w\|_{6}^{1/2}\,.
\ee
\end{lemma}
\par\smallskip\noindent
\textbf{Proof\,:} 
\beq{AppAp1}
\I |\bv_{z}|^{3}dV &=& \I (|u_{z}|^{2} + |v_{z}|^{2})^{3/2}dV\non\\
&\leq& \frac{3}{2} \I (|u_{z}|^{3} + |v_{z}|^{3})dV
\eeq
and, given the boundary conditions on $u$ and $v$, 
\beq{AppAp2}
\I |u_{z}|^{3}dV &=& \I u_{z}u_{z}|u_{z}|dV\non\\
&=& -\I \left\{u u_{zz}|u_{z}| + u u_{z} \frac{d |u_{z}|}{dz}\right\}dV + 
\int_{S^{\pm}}u u_{z} |u_{z}|dxdy\non\\
&\leq& 2\I|u| |u_{zz}||u_{z}|dV\,.
\eeq
which holds because $d|f|/dz \leq |f_{z}|$ for any appropriately function 
differentiable $f$. Thus (\ref{AppAp1}) becomes 
\beq{AppAp4}
\I |\bv_{z}|^{3}dV &\leq& 3 \I\left\{|u| |u_{zz}||u_{z}| + |v| |v_{zz}||v_{z}|\right\}dV\non\\
&\leq& 6\I |\bv| |\bv_{zz}||\bv_{z}| dV\non\\
&\leq& 6\|\bv\|_{6}\|\bv_{zz}\|_{2}\|\bv_{z}\|_{3}\,,
\eeq
which gives the advertised result. The result for $w$ follows in a similar manner. \hfill $\square$  
\par\medskip
Continuing with (\ref{newtemp5}), multiplying by $\Rey^{\gamma}$, where $\gamma$ is 
to be determined, (\ref{newtemp5}) becomes
\beq{newtemp7}
\shalf \frac{d~}{dt}\Rey^{\gamma}\I |\Theta_{z}|^{2}dV 
&\leq& -(\sigma\Rey)^{-1}\Rey^{\gamma}\I |\nabla_{3} \Theta_{z}|^{2}dV + 2^{1/2}
\varep\Rey^{\gamma}\|\Theta\|_{6}\|w\|_{6}^{1/2}\|w_{zz}\|_{2}^{1/2}\|\Theta_{zz}\|_{2}\non\\ 
&+& 6^{1/2}\Rey^{\gamma}\|\bv_{zz}\|_{2}^{1/2}\|\bv\|_{6}^{1/2} \|\Theta\|_{6}\left\{\|\Theta_{xz}\|_{2} 
+ \|\Theta_{yz}\|_{2}\right\}\non\\ 
&+& \Rey^{\gamma}\|\Theta_{zz}\|_{2}\|q\|_{2}\,.
\eeq
In turn, this re-arranges to
\beq{newtemp8}
\shalf\Rey^{\gamma}\frac{d~}{dt}\I |\Theta_{z}|^{2}dV 
&\leq& -(\sigma\Rey)^{-1}\Rey^{\gamma}\I |\nabla_{3}\Theta_{z}|^{2}dV\non\\
&+& \left[4\delta_{3}\Rey^{-1}\varep^{2}\|w_{zz}\|_{2}^{2}\right]^{1/4}
\left[\delta_{3}(\sigma\Rey)^{-1}\Rey^{\gamma}\|\Theta_{zz}\|_{2}^{2}\right]^{1/2}\non\\
&\times&\left[\delta_{3}^{-3}\sigma^{2}\Rey^{b_{1}}\|\Theta\|_{6}^{6}\right]^{1/6}
\left[\delta_{3}^{-3}\varep^{6}\sigma^{2}\Rey^{b_{2}}\|w\|_{6}^{6}\right]^{1/12}\non\\
&+& \big\{\left[4\delta_{4}\Rey^{-1}\|\bv_{zz}\|_{2}^{2}\right]^{1/4}
\left[3^{6}\delta_{4}^{-3}\Rey^{c_{2}}\|\bv\|_{6}^{6}\right]^{1/12}\big\}\non\\
&\times& \left[\frac{\sigma^{3}}{8\delta_{4}^{3}}\Rey^{c_{1}}\|\Theta\|_{6}^{6}\right]^{1/6}\,
\big\{2\delta_{4}(\sigma\Rey)^{-1}\Rey^{\gamma}\left[\|\Theta_{xz}\|_{2}^{2} 
+ \|\Theta_{yz}\|_{2}^{2}\right]\big\}^{1/2}\non\\
&+& \left\{\delta_{5}(\sigma\Rey)^{-1}\Rey^{\gamma}\|\Theta_{zz}\|_{2}^{2}\right\}^{1/2}
\left\{\delta_{5}^{-1}(\sigma\Rey)\Rey^{\gamma}\|q\|_{2}^{2}\right\}^{1/2}\,.
\eeq
where $2b_{1} + b_{2} = 9 + 6\gamma$ and $2c_{1} + c_{2} = 9 + 6\gamma$. Using Young's 
inequality it is found that
\beq{newtemp9}
\shalf\Rey^{\gamma}\frac{d~}{dt}\I |\Theta_{z}|^{2}dV 
&\leq& -(\sigma\Rey)^{-1}\Rey^{\gamma}\I|\nabla_{3}\Theta_{z}|^{2}dV\non\\
&+& \delta_{3}\Rey^{-1}\varep^{2}\|w_{zz}\|_{2}^{2}
+ \shalf\delta_{3}(\sigma\Rey)^{-1}\Rey^{\gamma}\|\Theta_{zz}\|_{2}^{2}\non\\
&+& \frac{1}{6}\delta_{3}^{-3}\sigma^{2}\Rey^{b_{1}}\|\Theta\|_{6}^{6}
+ \frac{1}{12}\delta_{3}^{-3}\varep^{6}\sigma^{2}\Rey^{b_{2}}\|w\|_{6}^{6}\non\\
&+& \delta_{4}\Rey^{-1}\|\bv_{zz}\|_{2}^{2} 
+ \delta_{4}(\sigma\Rey)^{-1}\Rey^{\gamma}\left[\|\Theta_{xz}\|_{2}^{2} 
+ \|\Theta_{yz}\|_{2}^{2}\right]\non\\
&+& \frac{1}{48\delta_{4}^{3}}\sigma^{3}\Rey^{c_{1}}\|\Theta\|_{6}^{6} + 
\frac{3^6}{12\delta_{4}^{3}}\Rey^{c_{2}}\|\bv\|_{6}^{6}\non\\
&+& \shalf \delta_{5}(\sigma\Rey)^{-1}\Rey^{\gamma}\|\Theta_{zz}\|_{2}^{2} + 
\shalf \delta_{5}^{-1}(\sigma\Rey)\Rey^{\gamma}\|q\|_{2}^{2}
\eeq
Gathering terms we find
\beq{newtemp10}
\shalf\Rey^{\gamma}\frac{d~}{dt}\I |\Theta_{z}|^{2}dV &\leq& -(\sigma\Rey)^{-1}\Rey^{\gamma}
\I \left\{(1-\delta_{4})(|\Theta_{xz}|^{2} + |\Theta_{yz}|^{2}) 
+ (1-\shalf\delta_{3}-\shalf\delta_{5})|\Theta_{zz}|^{2}\right\}dV\non\\
&+& \Rey^{-1}\left\{\delta_{4}\|\bv_{zz}\|_{2}^{2} + \delta_{3}\varep^{2}\|w_{zz}\|_{2}^{2}\right\}
+ \frac{3^6}{12\delta_{4}^{3}}\Rey^{c_{2}}\|\bv\|_{6}^{6}\non\\
&+& \left\{\frac{\sigma^{2}}{6\delta_{3}^{3}}\Rey^{b_{1}} 
+ \frac{\sigma^{3}}{48\delta_{4}^{3}}\Rey^{c_{1}}\right\}\|\Theta\|_{6}^{6}
+ \frac{\varep^{6}\sigma^{2}}{12\delta_{3}^{3}}\Rey^{b_{2}}\|w\|_{6}^{6}\non\\
&+& \shalf \delta_{5}^{-1}(\sigma\Rey)\Rey^{\gamma}\|q\|_{2}^{2}\,.
\eeq

\subsubsection{\sf A combination of the fluid and temperature inequalities}

(\ref{newtemp10}) is now combined with (\ref{comb2}) 
\beq{ft2}
\shalf\frac{d~}{dt}\I |\bvom|^{2}dV &+& \shalf\Rey^{\gamma}\frac{d~}{dt}\I |\Theta_{z}|^{2}dV\non\\
&\leq& -\left(1 - \threequart\delta_{1} - \delta_{2}\right)\Rey^{-1}\I|\hbox{curl}\,\bvom|^{2}dV\non\\
&-&(\sigma\Rey)^{-1}\Rey^{\gamma}\I \left\{(1-\delta_{4})(|\Theta_{xz}|^{2} + |\Theta_{yz}|^{2}) + 
(1-\shalf\delta_{3}-\shalf\delta_{5})|\Theta_{zz}|^{2}\right\}dV\non\\
&+& \Rey^{-1}\left\{\delta_{4}\|\bv_{zz}\|_{2}^{2} + \delta_{3}\varep^{2}\|w_{zz}\|_{2}^{2}\right)
+ \left(\frac{9}{4\delta_{1}^{3}}\Rey^{3} + \frac{3^6}{12\delta_{4}^{3}}\Rey^{c_{2}}\right)\|\bv\|_{6}^{6}\non\\ 
&+& \varep^{6}\left(\frac{3}{2\delta_{1}^{3}}\Rey^{3} + \frac{\sigma^{2}}{12\delta_{3}^{3}}\Rey^{b_{2}}\right)\|w\|_{6}^{6}
+ \left(\frac{\sigma^{2}}{6\delta_{3}^{3}}\Rey^{b_{1}} 
+ \frac{\sigma^{3}}{48\delta_{4}^{3}}\Rey^{c_{1}}\right)\|\Theta\|_{6}^{6}\non\\
&+& \frac{1}{\varep^{2}}\left(\delta_{2} + \frac{1}{2\delta_{2}}\right)\Rey\|\bv\|_{2}^{2} 
+ \frac{a_{0}^2}{2\varep^{2}\delta_{2}}\Rey\|\Theta\|_{2}^{2}
+ \shalf \delta_{5}^{-1}(\sigma\Rey)\Rey^{\gamma}\|q\|_{2}^{2}
\eeq
and 
\bel{param1}
\left.
\begin{array}{c}
2b_{1} + b_{2}\\
2c_{1} + c_{2}
\end{array}
\right\}
= 9 + 6\gamma
\ee
Our choices are
\beq{ch1}
b_{1} = c_{1} = -3,\qquad b_{2} = c_{2} = 3,\qquad\gamma = -2\,.
\eeq
$\delta_{3}$ and $\delta_{4}$ also need to be chosen such that the $\|\bv_{zz}\|^{2}$- 
and $\varep^{2}\|w_{zz}\|^{2}$-terms cancel from $\|\hbox{curl}\,\bvom\|_{2}^{2}$. 
To this end we choose $\delta_{0} = \shalf$ and make
\beq{delta0}
1 - \threequart\delta_{1} - \delta_{2} &=& \delta_{4}\non\\
\left(1 - \threequart\delta_{1} - \delta_{2}\right)(1-\delta_{0}) &=& \delta_{3}\,.
\eeq
Hence $\delta_{4} = 2\delta_{3}$. With this we choose $\delta_{5}$ such that the coefficients 
$1 - \delta_{4}$ and $1 - \shalf\delta_{3} - \shalf\delta_{5}$ within the double derivatives 
of the temperature are equal. Thus $\delta_{5} = 3\delta_{3}$. Together we have
\bel{ch2}
\delta_{3} = \shalf\big(1 - \threequart\delta_{1} - \delta_{2}\big),\qquad
\delta_{4} = 2\delta_{3},\qquad
\delta_{5} = 3\delta_{3}\,.
\ee
where $\delta_{1} > 0$ and $\delta_{2} > 0$ are arbitrarily chosen under the constraint 
that $\delta_{3}>0$.
\par\medskip\noindent 
Now we turn to the last three steps in the calculation\,:
\par\medskip\noindent
\textbf{Step 1\,:} To deal with the first set of terms on the right hand side of (\ref{ft2}) we use 
the expression for $\|\hbox{curl}\,\bvom\|_{2}^{2}$ in Lemma \ref{lemma1} with $\delta_{0}=\shalf$ 
and write 
\beq{ft4}
\I |\hbox{curl}\,\bvom|^{2}dV - \left(\|\bv_{zz}\|_{2}^{2} + \frac{1}{2}\varep^{2}\|w_{zz}\|_{2}^{2}\right)
> \I|\nabla_{2}\omega_{3}|^{2} + |u_{yz}|^{2} + |v_{xz}|^{2}\big\}dV
\eeq
This turns (\ref{ft2}) into
\beq{ft5}
\shalf\frac{d~}{dt}\I |\bvom|^{2}dV &+& \shalf\Rey^{-2}\frac{d~}{dt}\I |\Theta_{z}|^{2}dV\\
&\leq& - 2\delta_{3}\Rey^{-1}\I \big\{|\nabla_{2}\omega_{3}|^{2} + |u_{yz}|^{2} + |v_{xz}|^{2}\big\}dV\non\\
&-& \sigma^{-1}(1-2\delta_{3})\Rey^{-3}\I |\nabla_{3}\Theta_{z}|^{2}dV 
+ \left(\frac{9}{4\delta_{1}^{3}} + \frac{243}{32\delta_{3}^{3}}\right)\Rey^{3}\|\bv\|_{6}^{6}\non\\
&+& \varep^{6}\left(\frac{3}{2\delta_{1}^{3}} + \frac{\sigma^{2}}{12\delta_{3}^{3}}\right)\Rey^{3}\|w\|_{6}^{6}
+ \frac{\sigma^2}{6\delta_{3}^{3}}\left(1 + \frac{\sigma}{64}\right)\Rey^{-3}\|\Theta\|_{6}^{6}\non\\
&+& \varep^{-2}\Rey\left[\left(\delta_{2} + \frac{1}{2\delta_{2}}\right)
\|\bv\|_{2}^{2} + \frac{a_{0}^{2}}{2\delta_{2}}\|\Theta\|_{2}^{2}\right] + 
\frac{\sigma}{6\delta_{3}}\Rey^{-1}\|q\|_{2}^{2}\,.
\eeq
\par\medskip\noindent
\textbf{Step 2\,:} Now this inequality is re-scaled back to dimensional variables defined in 
Table \ref{table2} and $\mathcal{H}(t)$ defined in (\ref{Hdef}). 
This involves multiplying both sides of (\ref{ft5}) by $\Rey^3$. 
\beq{ft6}
\shalf\frac{d\mathcal{H}}{dt}
&\leq& \!-\frac{2\delta_{3}L^{2}}{\omega_{0}^{2}}\I\left(|\nabla_{2}\Omega_{3}|^{2} 
+ \asp^{2}|u_{1,x_{2}x_{3}}|^{2} + \asp^{2}|u_{2,x_{1}x_{3}}|^{2}\right)dV
- \frac{(1-2\delta_{3})L^{4}}{\sigma T_{0}^{2}}\I |\nabla T_{x_{3}}|^{2}dV\non\\
&+& \left(\frac{9}{4\delta_{1}^{3}} + \frac{243}{32\delta_{3}^{3}}\right)\|\mathcal{R}_{u_{hor}}\|_{6}^{6} + 
\left(\frac{3}{2\delta_{1}^{3}} + \frac{\sigma^{2}}{12\delta_{3}^{3}}\right)\|\mathcal{R}_{u_{3}}\|_{6}^{6}
+ \frac{\asp^{-6}\sigma^2}{6\delta_{3}^{3}}\left(1 + \frac{\sigma}{64}\right)\|\mathcal{R}_{a,T}\|_{6}^{6}\non\\
&+&  \varep^{-2}\left\{\left(\delta_{2} + \frac{1}{2\delta_{2}}\right)\Rey^{2}\|\mathcal{R}_{u_{1}}\|_{2}^{2} 
+ \frac{a_{0}^{2}\asp^{-6}}{2\delta_{2}}\Rey^{4}\|
\mathcal{R}_{a,T}\|_{2}^{2}\right\} + \frac{\sigma}{6\delta_{3}}\Rey\|q\|_{2}^{2}\,.
\eeq
\par\medskip\noindent
\textbf{Step 3\,:} Finally, we take the time integral over an interval $[0,\,\tau_{*}]$. 
\beq{ft7}
\int_{0}^{\tau_{*}}\I\big\{\!\!&-&\!\!\frac{2\delta_{3}L^{2}}{\omega_{0}^{2}}\big\{
|\nabla_{2}\Omega_{3}|^{2} + \asp^{2}|u_{1,x_{2}x_{3}}|^{2} + \asp^{2}|u_{2,x_{1}x_{3}}|^{2}\big\}
- \frac{(1-2\delta_{3})L^{4}}{\sigma T_{0}^{2}}|\nabla T_{x_{3}}|^{2}\non\\ 
&+& \beta_{u_{hor}}|\mathcal{R}_{u_{hor}}|^{6} + \beta_{u_{3}}|\mathcal{R}_{u_{3}}|^{6} + 
\beta_{1,T}|\mathcal{R}_{a,T}|^{6} + \beta_{2,T}\Rey^{4}\mathcal{R}_{a,T}^{2} + F_{H}(Q)\non\\ 
&+& \shalf \mathcal{V}_{*}^{-1}\mathcal{H}(0)\big \}dVdt \geq \shalf\mathcal{H}(\tau_{*})\,.
\eeq
Because of regularity \cite{CT05,CT07,Kob06,Kob07a} $\mathcal{H}(\tau_{*})$ is always under control 
from above and it also has a uniform lower bound $\mathcal{H}(\tau_{*}) > 0$ although zero may be a 
poor lower bound.
\par\medskip
Together with the use of Lemma \ref{lemma1} with $\delta_{0} = \shalf$, gives the result of Theorem 
\ref{hydrothm}, where $\beta_{u_{hor}},~\beta_{u_{3}}$ and $\beta_{i,T}$ are defined in Table \ref{table3} 
and the forcing function $F_{H}(Q)$ is defined in (\ref{Fdef}). \hfill $\blacksquare$

\section{\sf\textbf{Potential implications for simulations}}\label{implic}

The main result of this paper is that solutions of HPE can potentially develop extremely small scales 
of motion, allowed by the estimates derived here. These size scales decrease as $\mathcal{R}_{u}^{-1}$ 
and $\mathcal{R}_{a,T}^{-1}$, which means they could easily become of the order of metres or less at 
the very large values of these parameters achieved  in both atmospheric and oceanic flows. The 
hydrostatic estimate for the length scale defined $\lambda_{H}$ in (\ref{lsH}) is of the order of 
a \textit{metre} or less. Of course, this very small estimate may not be the thickness of a front\,; 
instead, it may refer to the smallest scale of features \textit{within} a front. The importance of the 
tendency to produce vigorous intermittent small scales in NWP and ocean circulation simulations remains 
to be determined but it may effect parameterizations as numerical resolution improves. In particular, 
one may ask whether parameterizations developed at coarser scales will still 
be accurate at finer scales, if the finer scales undergo the extreme events whose potential appearance
has been predicted in this paper. As for the perennial question of initial conditions, one must hope 
that flow activity initialized at coarse scales will be consistently followed to smaller scales without 
undue amplification of simulation errors. 

The consequences of Theorem 1 in \S \ref{events} is that space-time is potentially divided into two regions 
$\mathbb{S}^{+}$ and $\mathbb{S}^{-}$. The region $\mathbb{S}^{-}$ could be a union of a large number of 
disjoint sets, and if it were non-empty the flows in $\mathbb{S}^{-}$ would be dominated by strong concentrated 
structures. Very large lower bounds on double mixed derivatives of components of the velocity field 
$(u_{1},\,u_{2},\,u_{3})$ such as $|\partial^{2}u_{1}/\partial x_{3}\partial x_{2}|^{2}$ or 
$|\partial^{2}u_{2}/\partial x_{1}\partial x_{3}|^{2}$ may occur within $\mathbb{S}^{-}$, thus 
representing intense accumulation in the $(x_{1},x_{3})$- and $(x_{2},x_{3})$-planes respectively. 
For a nonempty $\mathbb{S}^{-}$, one would see the spontaneously formation of front-like objects 
localised in space that would only exist for a finite time. $\mathcal{R}_{u_{hor}}$ is a \textit{local} 
horizontal Reynolds number depending upon the local space-time values of $u(x_{1},x_{2},x_{3},\tau) 
= \sqrt{u_{1}^{2} + u_{2}^{2}}$ and $\mathcal{R}_{a,T}$ is a Rayleigh number dependent on the local temperature 
$T(x_{1},x_{2},x_{3},\tau)$. The large lower bounds on double-derivatives of solutions within the $\mathbb{S}^{-}$ 
regions can be converted into the large lower bounds on inverse length scales $\lambda_{H}^{-1}$. Thus to resolve 
a region such as this would require\footnote{\sf $\mathcal{R}_{u_{3}}$ is expected to  be negligible compared to 
$\mathcal{R}_{u_{hor}}$ because $u_{3}\sim O(\varepsilon)$.}
\bel{grd1}
\hbox{Number~of~grid~points} > const\,\big(\mathcal{R}_{u_{hor}}^{3} 
+ \mathcal{R}_{a,T}^{3} + \Rey^{2}\mathcal{R}_{a,T}\big)\,.
\ee
It is also worth remarking that the $L^6$-norm arising in the proof of Theorem \ref{hydroproof}, leading to the 
sixth powers of the \textit{local} Reynolds numbers $\mathcal{R}_{u_{hor}}$ for $\lambda_{H}^{-1}$, is precisely 
the norm that was proved by Cao and Titi \cite{CT05,CT07} to be bounded for HPE. While $\mathcal{R}_{u_{hor}}$ 
is a function of space-time it is a bounded function, but how much $\mathcal{R}_{u_{hor}}$ oscillates around its 
global space-time average $\Rey$ is unknown\,: this could 
vary significantly in different parts of the flow. Thus, how $\bu_{3} = \{u_{1},u_{2},\,u_{3}\}$ varies across a 
front is an important issue.  The limitations of the result are that no further information is available from 
the analysis regarding the spatial or temporal statistics of the subsets of $\mathbb{S}^{-}$ on which 
intense events would occur. 

If the regularity problem were to be settled in the NPE case, the results would likely be qualitatively the same
but with a non-negligible $\mathcal{R}_{u_{3}}$ term whose contribution may be significant in regions of strong 
vertical convection. There would also have to be significant technical differences\,: the domain would need to 
made periodic in the velocity variables and their derivatives because of lack of specification of horizontal 
velocity derivatives. 

Future improvements in numerical capabilities for the prediction of weather, climate and ocean 
circulation may be expected to enhance spatial and temporal resolutions. In addition, they will 
raise the issue of the optimal allocation of numerical resources. For example, improving the 
computations for parameterizations of other currently unresolved physical processes (such as 
phase changes in cloud physics) may have effects that are at least as significant as computing 
non-hydrostatic effects at finer resolution. Improvements in resolution will also raise the 
issue of whether subgrid-scale parameterizations of these unresolved physical processes that 
have been developed for numerical prediction at coarser scales will transfer accurately to 
computations at finer scales, regardless of whether the hydrostatic approximation is retained. 
Thus, one may expect the HPE to remain central in the discussions about choices among the 
various potential numerical code implementations for weather, climate and ocean circulation 
predictions well into the foreseeable future. Even though they are mathematically well-posed, 
the HPE have been shown here to contain the potential for sudden, localized events to occur on 
extremely small scales in space and time. 

\par\medskip\noindent
\textbf{Acknowledgements\,:} We thank Peter Bartello, Raymond Hide, Brian Hoskins, Tim Palmer 
\& Edriss Titi for several enlightening conversations. Darryl Holm thanks the Royal Society 
for a Wolfson Research Merit Award.

\rem{
\vspace{2cm}
Now
\beq{z1}
\bV\cdot\nabla_{3}\bv &=& \bV\cdot\nabla_{3}\bV - \varep\bk\bV\cdot\nabla_{3}w\non\\
\bV\cdot\nabla_{3}\bV &=& \shalf \nabla_{3}(u_{3}^{2}) - \bV\times\bom
\eeq
Therefore
\beq{z2}
\hbox{curl}\,(\bV\cdot\nabla_{3}\,\bv) &=& \hbox{curl}\,(\bom\times\bV) -
\varep\hbox{curl}\,(\bk\bV\cdot\nabla_{3}w)\non\\
&=& \bV\cdot\nabla_{3}\bom - \bom\cdot\nabla_{3}\bV -
\varep \hbox{curl}\,(\bk\bV\cdot\nabla_{3}w)\non\\
&=& \bV\cdot\nabla_{3}\left\{\bvom - \varep \nabla^{\perp}w\right\}
- \left\{\bvom - \varep\nabla^{\perp}w\right\}\cdot\nabla_{3}\bV
- \varep \hbox{curl}\,(\bk\bV\cdot\nabla_{3}w)\non\\
&=& \bV\cdot\nabla_{3}\bvom - \bvom\cdot\nabla_{3}\bv - \varep\bk\bvom\cdot\nabla_{3}w 
-\varep \bV\cdot\nabla_{3}(\nabla^{\perp}w)\non\\
&+& \varep (\nabla^{\perp}w)\cdot\nabla_{3}\bV
+ \varep\nabla^{\perp}(\bV\cdot\nabla_{3}w)
\eeq
because $\hbox{curl}\,(\bk A) = -\nabla^{\perp}A$\,. We consider the group of the last three 
terms (without the common $\varep$-coefficient)
\beq{z3}
&-&\bV\cdot\nabla_{3}(\nabla^{\perp}w) + (\nabla^{\perp}w)\cdot\nabla_{3}\bV
+ \nabla^{\perp}(\bV\cdot\nabla_{3}w)\non\\
&=&  (\nabla^{\perp}w)\cdot\nabla_{3}\bV +  (\nabla^{\perp}\bV)\cdot\nabla_{3}w\non\\
&=& \left(-w_{y}\partial_{x} + w_{x}\partial_{y}\right) \bV -\bi \bu_{3,y}\cdot\nabla_{3}w
+ \bjvec \bu_{3,x}\cdot\nabla_{3}w\non\\
&=& -\bivec w_{y}\left(u_{x}+v_{y} + \varep w_{z}\right) + \bjvec w_{x}\left(u_{x}+v_{y} 
+ \varep w_{z}\right)\non\\
&=& 0\,,
\eeq
where we have invoked $\hbox{div}\,\bV = 0$ on the last line. From (\ref{z2}) we are 
finally left with 
\beq{z4}
\hbox{curl}\,(\bV\cdot\nabla_{3}\,\bv) &=& 
\bV\cdot\nabla_{3}\bvom - \bvom\cdot\nabla_{3}\bv - \varep\bk\bvom\cdot\nabla_{3}w\non\\
&=& \bV\cdot\nabla_{3}\bvom - \bvom\cdot\nabla_{3}\bV\,.
\eeq
We have finally shown that $\bvom$ satisfies
\bel{z5}
\varep \frac{D\bvom}{Dt} = \varep \bvom\cdot\nabla_{3}\bV + \varep \Rey^{-1}\Delta_{3}\bvom + 
\hbox{curl}\,(\bk\times\bv + a_{0}\bk T) \,.
\ee
}
\rem{
\newpage
\section{What is $a_{0}$?}
\label{app1}

Here we take straight 3D-NS coupled to a temperature, but without Richardson's scalings, 
to see what the coefficient $a_{0}$ in the hydrostatic equation should be. Consider the 
original equations in primed dimensional variables
\bel{ap1}
\bu'_{\tau} +\bu'\cdot\nabla'\bu' + f(\bk\times\bu') = \nu\Delta' \bu' -\nabla' p' + \alpha g T\,\bk
\ee
Re-scale this into unprimed, dimensionless variables\,: $\bu = U_{0}^{-1}\bu',~\bx = L^{-1}\bx_{1},
~t = U_{0}L^{-1}\tau$ and $p = U_{0}^{-1}(Lf)^{-1}p'$ to give
\bel{ap2}
U_{0}^{2}L^{-1}\left(\bu_{t} + \bu\cdot\nabla\bu\right) + U_{0}f(\bk\times\bu)
= U_{0}L^{-2}\nu\Delta \bu - U_{0}f\nabla p + \alpha g T\,\bk
\ee
Therefore, with $\Rey = LU_{0}\nu^{-1}$ and $\varep = U_{0}/Lf$, we find
\bel{ap3}
\bu_{t} + \bu\cdot\nabla\bu + \varep^{-1}(\bk\times\bu) = \Rey^{-1}\Delta \bu - \varep^{-1}\nabla p + 
\left(\frac{LT_{0}\alpha g}{U_{0}^2}\right)TT_{0}^{-1}\,\bk
\ee
Given that the Rayleigh and Prandtl numbers and the aspect ratio are defined by
\bel{ap4}
R_{a} = \frac{\alpha g H^{3}T_{0}}{\nu\kappa}\qquad\qquad
\sigma = \frac{\nu}{\kappa}\qquad\qquad B = \frac{H}{L}
\ee
we find that 
\bel{a0defex}
a_{0} = \frac{LT_{0}\alpha g}{U_{0}^2} = \frac{\asp^{-3}R_{a}}{\sigma\Rey^{2}}\,.
\ee
with 
\bel{ap6}
\bu_{t} + \bu\cdot\nabla\bu + \varep^{-1}(\bk\times\bu) 
= \Rey^{-1}\Delta \bu - \varep^{-1}\nabla p + a_{0}TT_{0}^{-1}\,\bk
\ee
}



\begin{thebibliography}{14}\itemsep -1.5mm
\small

\bibitem{Rich22} Richardson, L. F. 1922 \textit{Weather Prediction by Numerical Process} 
(Cambridge: Cambridge University Press) reprinted 1988 by New York: Dover.

\bibitem{Charn55} Charney, J. G. 1955 The use of the primitive equations of motion in 
numerical prediction, \textit{Tellus}, \textbf{7}, 22--26.

\bibitem{Eck60} Eckart, C. 1960 \textit{The Hydrodynamics of Oceans and Atmospheres}, 
Oxford\,: Pergamon Press.

\bibitem{Gill82} Gill, A. 1982 \textit{Atmosphere-ocean dynamics}, London: Academic Press.

\bibitem{Ped87} Pedlosky, J, 1987 \textit{Geophysical Fluid Dynamics}, 2nd edition 
New York: Springer-Verlag.

\bibitem{Salmon88} Salmon, R. 1988 Hamiltonian fluid mechanics \textit{Ann. Rev. Fluid Mech.}, 
\textbf{20}, 225--256.

\bibitem{Ho92} Holton, J. R. 1992 \textit{An Introduction to Dynamic Meteorology} 3rd edition, 
San Diego: Academic Press.

\bibitem{Cullen93} Cullen, M. J. P. 1993 The unified forecast/climate model,\textit{Meterol. 
Mag.}, \textbf{122},81--94.

\bibitem{Holm96} Holm, D. D. 1996 Hamiltonian balance equations,\textit{Physica D},textbf{98},379--414.

\bibitem{Marshall97} Marshall, J., Hill, C., Perelman, L. and Adcroft, A. 1997 Hydrostatic, quasi-hydrostatic
and non-hydrostatic ocean modelling, \textit{J. Geophys. Res.}, \textbf{102}, 5733--5752.

\bibitem{NR02} Norbury, J. and Roulstone, I. 2002 \textit{Large-scale atmosphere-ocean dynamics I \& II},
Cambridge: Cambridge University Press.

\bibitem{White02} White, A. A. 2002 A view of the equations of meteorological dynamics and 
various approximations, in Norbury, J. and Roulstone, I. (eds) \textit{Large-scale 
ocean-atmosphere dynamics I}, Cambridge: University Press.

\bibitem{White03}  White, A. A. 2003 Primitive Equations in Holton J. R. \textit{et al} (eds) 
\textit{Encyclopedia of atmospheric science} pp 694--702, New York: Academic Press.  

\bibitem{DaCuMaMaStWhWo05} Davies, T., Cullen, M. J. P., Malcolm, A. J., Mawson M,., H., Staniforth, A., 
White, A. A. and Wood N. 2005 A new dynamical core for the Met Office's global and regional 
modelling of the atmosphere, \textit{J. R. Met. Soc.} \textbf{131}, 1759--1782.

\bibitem{WhHoRoSt05} White, A. A., Hoskins, B. J., Roulstone, I. and Staniforth, A. 2005 Consistent 
approximate models of the global atmosphere: shallow, deep, hydrostatic, quasi-hydrostatic 
and non-hydrostatic, \textit{Quart. J. R. Met. Soc.} \textbf{131}, 2081--2107.

\bibitem{Cullen06} Cullen, M. J. P. 2006 \textit{A Mathematical Theory of Large-scale 
Atmosphere/Ocean Flow}, London: Imperial College Press.

\bibitem{Cullen07} Cullen, M. J. P. 2007 Modelling Atmospheric Flows, \textit{Acta Numerica}, 1--87.

\bibitem{WS09} Wedi, N. P. and Smolarkiewicz, P. K. 2009 A framework for testing global non-hydrostatic
models, \textit{Q. J. R. Meteorol. Soc.} (Published online:DOI: 10.1002/qj)

\bibitem{OhSaMaNa05} Ohfuchi, W., Sasaki. H., Masumoto, Y. and Nakamura, H. 2005 Mesoscale 
resolving simulations of the global atmosphere and ocean on the Earth Simulator 
\textit{EOS Trans. AGU}, \textbf{86}, 45--46.

\bibitem{SACRLLC06} Shen, B.-W., Atlas, R., Chern, J.-D., Reale, O., Lin, S.-J., Lee, T. and Chang, J. 
2006 The 0.125 degree finite-volume general circulation model on the NASA Columbia supercomputer: 
preliminary simulations of mesoscale vortices, \textit{Geophys. Res. Lett.}, \textbf{33}, L05801.

\bibitem{Ham08} Hamilton, K., Takahashi Y. O., and Ohfuchi, W. 2008 The mesoscale spectrum 
of atmospheric motions investigated in a very fine resolution global general circulation 
model, \textit{J. Geophys. Res.}, \textbf{113}, D18110. doi:10.1029/2008JD009785

\bibitem{El48} Eliassen, A. 1948 The quasi-static equations of motion, \textit{Geofys Publikasjoner}, 
\textbf{17}, No. 3.

\bibitem{H75} Hoskins, B. J. 1975 The Geostrophic momentum approximation and the 
semi-geostrophic equations, \textit{J. Atmos. Sci.}, \textbf{32}, 233--242.

\bibitem{Hrev82} Hoskins, B. J. 1982 The mathematical theory of frontogenesis, \textit{Ann. 
Rev. Fluid Mech.} \textbf{14}, 131--151.

\bibitem{HB72} Hoskins, B. J. and Bretherton, F. 1972 Atmospheric frontogenesis models; Mathematical 
formulation and solution, \textit{J. Atmos. Sci.}, \textbf{29}, 11--37. 

\bibitem{Shapiro84} Shapiro, M. A. 1984 Meteorological tower measurements of a surface cold front
\textit{Mon. Weather Rev.}, \textbf{112}, 1634-1639.

\bibitem{gust1} Weaver, J. F. and Purdom, J. F. W. 1995 An Interesting Mesoscale Storm-Environment 
Interaction Observed Just Prior to Changes in Severe Storm Behavior, \textit{Weather and Forecasting}, 
\textbf{10}, Issue 2, 449–-453.  


\bibitem{CT05} Cao, C. and Titi, E. S. 2005 \textit{Global well-posedness of the three-dimensional 
primitive equations of large scale ocean and atmosphere dynamics} arXiv: Math. AP/0503028

\bibitem{CT07} Cao, C. and Titi, E. S. 2007 Global well-posedness of the three-dimensional primitive 
equations of large scale ocean and atmosphere dynamics, \textit{Ann. Math.} \textbf{166}, 245–-267. 

\bibitem{Kob06} Kobelkov, G. 2006 Existence of a solution ``in whole'' for the large-scale 
ocean dynamics equations, \textit{Comptes Rendus Acad. Sci. Paris I}, \textbf{343}, 283–-286.

\bibitem{Kob07a} Kobelkov, G. 2007 Existence of a Solution ``in the large'' for Ocean Dynamics
Equations, \textit{J. Math. Fluid Mech.} \textbf{9}, no. 4, 588--610.

\bibitem{Kob07b} Kobelkov, G., 2008 Existence and uniqueness of a solution to primitive equations 
with stratification in the large, \textit{Russian J. Numer. Anal. Math. Modelling}, \textbf{23}, 39--61.

\bibitem{NJ07} Ju, N. 2007 The global attractor for the solutions of the $3D$ viscous 
Primitive Equations, \textit{Disc. Cont. Dyn. Systems}, \textbf{17}, 159--179. 

\bibitem{thin1} Raugel, G. and Sell, G. R. 1993 Navier-Stokes equations on thin 3D domains I\,; 
Global attractors and global regularity of solutions, \textit{J. Amer. Math. Soc.}, \textbf{6},  503-–568.

\bibitem{thin2} Raugel, G. and Sell, G. R. 1994 Navier-Stokes equations on thin 3D domains II; 
Global regularity of spatially periodic solutions, \textit{Nonlinear partial differential 
equations and their applications}, Coll\`{e}ge de France Seminar Vol XI, pp 205–-247, 
Harlow: Longman Sci. Tech.

\bibitem{thin3} Hu, C., Temam, R. and Ziane, M. 2003 The primitive equations on the large scale ocean 
under the small depth hypothesis, \textit{Discrete Contin. Dynam. Systems}, \textbf{9}, 97--131.



\bibitem{LTW1} Lions, J.,  Temam, R. and Wang, S. 1992 New formulations of the primitive 
equations of atmosphere and applications, \textit{Nonlinearity}, \textbf{5}, 237--288.

\bibitem{LTW2} Lions, J.,  Temam, R. and Wang, S. 1992 On the equations of the large scale Ocean 
\textit{Nonlinearity}, \textbf{5}, 1007--1053.

\bibitem{LTW3} Lions, J., Temam, R. and Wang, S. 1995 Mathematical theory for the coupled 
atmosphere-ocean models, \textit{J. Math. Pures Appl.}, \textbf{74}, 105--163.

\bibitem{Lewan07} Lewandowski, R. 2007 R\'{e}sultat d'existence d'une solution faible
au syst\`{e}me des equations primitives, \textit{Analyse Math\'{e}matique et oc\'{e}anographie\,:
Essai sur la mod\'{e}lisation et l'analyse Math\'{e}matique de quelques mod\`{e}les 
de turbulence utilis\'{e}s en oc\'{e}anographie}, chapter 2, June 15.

\bibitem{JDG09} Gibbon, J. D. 2009 Estimating intermittency in three-dimensional 
Navier-Stokes turbulence, \textit{J. Fluid Mech.} \textbf{625}, 125-133.

\bibitem{RH1} Hide, R. 1964 The viscous boundary layer at the free surface of a rotating 
baroclinic fluid, \textit{Tellus}, \textbf{16}, 523--529.

\bibitem{RH2} Hide, R. 1965 The viscous boundary layer at the free surface of a rotating 
baroclinic fluid: effects due to the temperature dependence of surface tension, 
\textit{Tellus} \textbf{17}, 440--442.

\bibitem{RH3} Hide, R. 1969 Some laboratory experiments on free thermal convection in a rotating 
fluid subject to a horizontal temperature gradient and their relation to the global atmospheric 
circulation, pp 196-221 in \textit{The Global Circulation of the Atmosphere} (ed. Corby G A), 
London: Royal Meteorological Society.







\end{thebibliography}
\end{document}